\documentclass[aps,pre,reprint,notitlepage,superscriptaddress]{revtex4-2}

\usepackage{cmap} 
\usepackage[T1]{fontenc} 

\usepackage{amsmath,amsfonts}
\usepackage{mleftright} 
\usepackage{mathtools} 

\usepackage{lmodern,amssymb}

\let\what\widehat

\usepackage{dsfont} 

\usepackage{siunitx}
\usepackage{graphicx}
\graphicspath{{figures/}}
\newlength\figwidth
\usepackage{url,hyperref}
\usepackage[table,dvipsnames]{xcolor}
\hypersetup{colorlinks=true, linkcolor=BrickRed,
  urlcolor=blue!50!black, citecolor=blue!50!black}

\usepackage[capitalise]{cleveref}

\usepackage{enumitem}
\usepackage{booktabs,tabularx}

\renewcommand\epsilon{\varepsilon}
\renewcommand\phi{\varphi}
\renewcommand\theta{\vartheta}
\renewcommand\rho{\varrho}
\renewcommand\vec[1]{\boldsymbol #1}
\newcommand\unitvec[1]{\vec{#1}}
\newcommand\tens[1]{\boldsymbol{#1}}
\newcommand\diff{\mathrm{d}}

\newcommand\e{\text{e}}

\renewcommand\geq\geqslant
\renewcommand\leq\leqslant

\DeclareMathOperator\Var{Var}

\newcommand\kB{k_{\text{B}}}

\newcommand\DR{D_\text{R}}
\newcommand\tauR{\tau_\text{R}}
\newcommand\ext{\text{ext}}
\newcommand\Lu{\mathcal{L}_{\vec u}}

\begin{document}

\title{Langevin equations and a geometric integration scheme for the overdamped limit of rotational Brownian motion of axisymmetric particles}

\newcommand\FUBaffiliation{\affiliation{Freie Universit\"{a}t Berlin, Institute of Mathematics, Arnimallee 6, 14195 Berlin, Germany}}
\newcommand\ZIBaffiliation{\affiliation{Zuse Institute Berlin, Takustra{\ss}e 7, 14195 Berlin, Germany}}

\author{Felix H{\"o}f{}ling}%
\FUBaffiliation
\ZIBaffiliation

\author{Arthur V. Straube}%
\ZIBaffiliation
\FUBaffiliation

\begin{abstract}
The translational motion of anisotropic or self-propelled colloidal particles is closely linked with the particle's orientation and its rotational Brownian motion.
In the overdamped limit, the stochastic evolution of the orientation vector follows a diffusion process on the unit sphere and is characterized by an orientation-dependent (``multiplicative'') noise.
As a consequence, the corresponding Langevin equation attains different forms depending on whether Itō's or Stratonovich's stochastic calculus is used.
We clarify that both forms are equivalent and derive them in a top-down appraoch from a geometric construction of Brownian motion on the unit sphere, based on infinitesimal random rotations.
Our approach suggests further a geometric integration scheme for rotational Brownian motion, which preserves the normalization constraint of the orientation vector exactly.
We show that a simple implementation of the scheme, based on Gaussian random rotations, converges weakly at order 1 of the integration time step,
and we outline an advanced variant of the scheme that is weakly exact for an arbitrarily large time step.
Due to a favorable prefactor of the discretization error, already the Gaussian scheme allows for integration time steps that are one order of magnitude larger compared to a commonly used algorithm for rotational Brownian dynamics simulations
based on projection on the constraining manifold.
For torques originating from constant external fields, we prove by virtue of the Fokker--Planck equation that the constructed diffusion process satisfies detailed balance and converges to the correct equilibrium distribution.
The analysis is restricted to time-homogeneous rotational Brownian motion (i.e., a single rotational diffusion constant), which is relevant for axisymmetric particles and also chemically anisotropic spheres, such as self-propelled Janus particles.
\end{abstract}

\maketitle

\section{Introduction}

The past two decades have seen considerable advances in the synthesis and characterization of anisotropic colloidal particles,
ranging from complex shapes, e.g., ellipsoids \cite{Han:2006,Mukhija:2007}, clusters \cite{Kraft:2013}, and propellers \cite{Schamel:JACS2013}, to anisotropic chemical surface coatings.
An important example for the latter are so-called Janus spheres \cite{Howse:2007,Erbe:JPCM2008,Palacci:PRL2010,Lee:2014,Buttinoni:JPCM2012,Volpe:SM2011},
which exhibit autonomous self-propulsion under suitable conditions and often serve as colloidal models for microswimmers.
The use of such particles in experimental model systems has opened new avenues and has boosted research in the field of active matter (see, e.g., Refs.~[\citenum{Bechinger:RMP2016,Palagi:AOM2019,Gibbs:L2019}]).
These successes have stimulated a large number of theoretical investigations, which have brought forth elaborate descriptions of the self-propelled motion of a single active colloid (see, e.g., Refs.~[\citenum{Schweitzer:PRL1998, Erdmann:EPJB2000, Popescu:2010, Romanczuk:EPJST2012, Sabass:2012, Colberg:2014}]).
Depending on the question at hand, hydrodynamic effects due to solvent flows can be crucial \cite{Pooley:PRL2007, Uspal:SM2015, Elgeti:RPP2015, Zoettl:ARCMP2023} or may be negligible.
In the latter case, the active Brownian particle (ABP) model is widely used
to describe the motion of active colloids \cite{Golestanian:2005,Romanczuk:EPJST2012,Bickmann:PRE2023}; it combines the Brownian motion of a colloidal Janus sphere with a constant propulsion velocity which randomly changes its direction.
The propulsion direction is described by a unit vector $\vec u$ which itself performs a Brownian motion on the unit sphere.
Thereby, the stochastic motion of $\vec u$ may evolve freely \cite{Howse:2007,tenHagen:2011,Choudhury:NJP2017} or under the influence of gravity \cite{tenHagen:NC2014, Wolff:EPJE2013, Ruehle:NJP2018, Vachier:EPJE2019, Brosseau:SM2021, Chepizhko:PRL2022}
and it becomes particularly challenging in the case of shape anisotropy \cite{Makino:2004, *Doi:2005, delaTorre:JNET2024, Wittkowski:2012, Needle_zigzag:2008, Needle_rotation:2008, Needle_Perrin:2009, Leitmann:2016,*Leitmann:2017, Kurzthaler:SR2016}.
The dynamics of active colloids becomes complex in the presence of rotation--translation coupling,
e.g., due to short-time hydrodynamic friction \cite{Makino:2004, *Doi:2005, delaTorre:JNET2024, Wittkowski:2012} or due to confinement by a harmonic potential well \cite{Caraglio:PRL2022} or a substrate potential \cite{Choudhury:NJP2017,Straube:PRE2024};
in such situations, the description of the rotational dynamics is crucial for predictions on the translational motion.

Models for the rotational motion of molecules have traditionally been studied in the context of dynamic light scattering and dielectric spectroscopy \cite{Coffey:LangevinEquation, BernePecora:DynamicLightScattering}.
These models are typically based on the Fokker--Planck equation for orientational diffusion \cite{Perrin:1934, Furry:1957, DaleFavro:1960}, which allows one to calculate correlation functions and, more generally, the statistics of the trajectories \cite{Needle_Perrin:2009, Leitmann:2016,*Leitmann:2017, Kurzthaler:SR2016}.
Alternatively, some form of Langevin equation is employed, which provides a description on the level of stochastic trajectories and which forms the basis of stochastic simulations.
The underdamped Langevin equation is used when inertia effects are relevant \cite{Coffey:LangevinEquation} and, mathematically, there are no conceptual questions with this equation apart from being demanding to work with it analytically.
In the context of colloidal motion, however, one would like to model the stochastic dynamics as (completely) overdamped, which requires a suitable Wiener process for the particle orientation.
The corresponding Langevin equation contains a noise term that has a ``multiplicative'' structure
\cite{Coffey:LangevinEquation, Makino:2004, *Doi:2005, delaTorre:JNET2024, Wittkowski:2012, Wolff:EPJE2013, Vachier:EPJE2019, Bickmann:PRE2023, Leitmann:2016,*Leitmann:2017, Kurzthaler:SR2016, Choudhury:NJP2017, Straube:PRE2024} [see below, \cref{eq:kinematic-Str,eq:kinematic-Ito}],
which brings up questions about the underlying stochastic calculus.
In particular, in its Itō form, the overdamped Langevin equation [\cref{eq:kinematic-Ito}] contains a drift-like term, which may appear counter-intuitive at first sight.
Another example for this mathematically well-understood situation \cite{Gardiner:Stochastic,Oksendal:Stochastic} is given by translational diffusion near a wall, characterized by a position-dependent diffusivity, which has lead to a re-arising apparent controversy over decades
\cite{Ryter:1981, Sancho:1982, Lau:PRE2007, Volpe:PRL2010, Sokolov:CP2010, Sancho:PRE2011, Pesce:NC2013, Farago:PRE2014, Volpe:RPP2016, Bhattacharyay:JPA2020}.
The issue of multiplicative noise in rotational diffusion is entirely circumvented by modeling the particle orientation as confined to a plane, which allows one to parametrise the two-dimensional rotations by a single angle.
Whereas this choice is widely adopted in models of active matter, it often appears to be a choice of convenience for simplicity rather than being rooted in physics. Notably, the linear transport behavior can sensitively depend on the dimensionality of rotational diffusion \cite{Straube:PRE2024}.

In this note, we aim to clarify some mathematical aspects of rotational Brownian motion, in particular, the different forms of the overdamped Langevin equation and how they depend on the adopted definition of the stochastic integral (Itō or Stratonovich).
The equations are derived from a geometric construction of Brownian motion on the unit sphere, given as a sequence of infinitesimal random rotations.
The construction suggests a geometric integration scheme for rotational Brownian motion, which (i) exactly preserves the normalization constraint of the orientation vector and (ii) can be extended such that it is weakly exact for arbitrarily large time steps, i.e., it correctly generates the complete probability distributions, including the propagator \cite{Gardiner:Stochastic,KloedenPlaten:NumericalSolution}.
(A strongly convergent scheme would require that the numerical solutions also agree at the level of the individual random trajectories.)

In order to focus on the essence of the problem, we shall mainly discuss chemically anisotropic spheres (e.g., Janus spheres), thereby avoiding additional difficulties that arise from the incessant transformation of the diffusivity tensor due to the rotational motion. We then show that the results apply also to axi\-symmetric particles, which requires some mostly technical modifications of the derivations.
In mathematical texts, such diffusion processes with constant coefficients are referred to as \emph{homogeneous in time}.

Finally, we mention that the literature describes a number of numerical schemes for Brownian motion of rigid bodies, which are often justified only heuristically;
this includes schemes using finite rotations \cite{Fernandes:2002,Beard:BJ2003}, adding a constraint force \cite{Ilie:2015,Tao:2005,Evensen:MTS2008,Waszkiewicz:SPPC2023},
discretizing the Stratonovich--Langevin equation \cite{Makino:2004,Gordon:PRE2009},
or employing a symplectic splitting scheme of the underdamped Langevin dynamics \cite{Sun:JCP2008}.
Also, the algorithm presented below (\cref{sec:algorithm}) has already been used in previous work of both authors \cite{Leitmann:2016,*Leitmann:2017, Choudhury:NJP2017, Straube:PRE2024}, without further explanation so far.

\section{Heuristic derivation of the overdamped limit}
\label{sec:heuristic}

\subsection{Langevin equation of rotational Brownian motion with inertia}

Let us start from classical mechanics and consider the rotational motion of a rigid body in three-dimensional space,
which may be subject to a total torque $\vec T(t)$ at time $t$.
Working in a space-fixed (or inertial) frame of reference, the body's angular velocity $\vec \omega(t)$, obeys Euler's equation, which is the rotational analogue of Newton's law:
\begin{equation}
  \frac{\diff}{\diff t} [\tens I(t) \, \vec \omega(t)] = \vec{T}(t) \,;
  \label{eq:euler}
\end{equation}
here, $\tens I(t)$ is the body's moment of inertia, which depends on its instantaneous orientation in space.
We consider a body-fixed axis, which is denoted by the unit vector $\vec u(t)$. Its time evolution given $\vec \omega(t)$ obeys the kinematic equation
\begin{equation}
  \dot{\vec u}(t) =  \vec\omega(t) \times \vec u(t) \,.
  \label{eq:kinematic}
\end{equation}
One readily verifies that this dynamics preserves the norm $|\vec u(t)| = 1$, since $\diff |\vec u|^2 /\diff t = 2 \vec u \cdot \dot{\vec u} = 0$.

If the particle is immersed in a fluid medium, it experiences a torque $-\tens \zeta_R(t) \, \vec \omega(t)$ due to viscous (Stokes) friction. In general, the rotational friction coefficient $\tens \zeta_R(t)$ is a tensor which is constant in a body-fixed frame  \cite{HappelBrenner:Hydrodynamics}, but is time-dependent in the space-fixed frame used here.
Following Langevin's and Ornstein's approach to Brownian motion \cite{vanKampen:StochasticProcesses}, the incessant collisions of the particle with solvent molecules are further subsumed in a random torque $\vec \xi(t)$ acting on the particle.
Taking the collisions to be uncorrelated in a first approximation suggests to model $\vec\xi(t)$ as a Gaussian white noise process, which is independent of $\vec \omega(s)$ for $t > s$, has mean zero, and its covariance matrix is determined by the fluctuation--dissipation relation:
\begin{equation}
  \langle \vec\xi(t) \rangle = 0 \quad \text{and} \quad
  \langle \vec\xi(t) \otimes \vec\xi(t') \rangle =2 \kB T \tens \zeta_R(t) \delta(t-t') \,,
  \label{eq:xi-moments}
\end{equation}
where $\kB T$ denotes the thermal energy scale
and $\langle \,\cdot\, \rangle$ is an ensemble average over the noise $\vec \xi(t)$.
Foreseeing the possibility of an externally imposed torque $\vec T_\ext(t)$, the total torque $\vec T$ has two deterministic and one random contributions so that \cref{eq:euler} is turned into the Euler--Langevin equation for the evolution of $\vec \omega(t)$:
\begin{equation}
  \frac{\diff}{\diff t} [\tens I(t) \, \vec \omega(t)]
  = \vec T_\ext(t) - \tens \zeta_R(t) \, \vec \omega(t) + \vec \xi(t) \,.
  \label{eq:euler-langevin}
\end{equation}
This is a stochastic differential equation (SDE) in $\vec \omega(t)$ and since the white noise $\vec \xi(t)$ enters ``additively'' (i.e., it is not multiplied by a $\vec \omega(t)$-dependent function), there is no ambiguity about the choice of the stochastic calculus:
any (consistent) definition of the stochastic integral yields the same solution \cite{Oksendal:Stochastic,Gardiner:Stochastic}.

The above assumptions on the random torque are very well fulfilled in many use cases. However, we remind the reader that
high-resolution measurements on colloidal particles have revealed deviations from \cref{eq:xi-moments} in the form of persistent correlations, due to hydrodynamic memory \cite{Franosch:2011,Huang:2011}.
Also at short lag times, time-reversibility and smoothness of the trajectories, at the scale of molecules and atoms, implies a vanishing high-frequency friction \cite{Straube:CP2020} and thus $\langle |\vec \xi(0)|^2 \rangle < \infty$, different from \cref{eq:xi-moments}.
A predictive theory of the emergence of friction from Hamiltonian dynamics remains as an open problem in statistical physics.
Nevertheless, one can phenomenologically account for both aspects in the framework of the generalized Langevin equation,
developed in the 1960's by Zwanzig \cite{Zwanzig:1965}, Mori \cite{Mori:1965a} and others, see Ref.~\citenum{Schilling:PR2022} for a recent review.
Integrating out the solvent degrees of freedom in rotational Brownian motion is a program that, to our knowledge, still needs to be carried out.
The use of a differentiable noise $\vec \xi(t)$ in \cref{eq:euler-langevin} would eliminate any ambiguity in the stochastic integral \cite{Kallenberg:Probability}, at the price of giving up the simplicity of a Markovian dynamics in Langevin's approach.

For the purpose of keeping the present discussion focused, we will limit ourselves to spherical particles with a chemically patterned surface so that their orientation remains identifiable; colloidal Janus particles are a prominent example.
In this case, the friction tensor is proportional to the unit matrix and, thus, becomes time-independent:
$\tens\zeta_R(t) = \zeta_R \mathds{1}$.
Correspondingly, the moment of inertia reduces to a constant scalar as well:
$\tens I(t) = I \mathds{1}$.
If the sphere has the mass $m$ and the radius $a$ and is suspended in a solvent of dynamic viscosity $\eta$ and density $\rho$,
one finds $I=(2/5)m a^2$ and the solution of the Stokes flow yields $\zeta_R= 8 \pi \eta a^3$ \cite{HappelBrenner:Hydrodynamics}.

In \cref{sec:axisymmetric}, we extend the results to axisymmetric particles, for which $\vec u(t)$ is naturally chosen to align with the symmetry axis of the particle. This axis fully determines the particle's orientation such that the friction tensor $\tens \zeta_R(t)$ and moment of inertia $\tens I(t)$ are functions of $\vec u(t)$.
For particles of arbitrary shape, the orientation may be tracked by following the motion of a body-fixed trihedron \cite{Makino:2004, *Doi:2005, delaTorre:JNET2024}, an elaboration on this generalization is left for future research.

\subsection{Heuristic derivation of the overdamped limit}

For small colloidal (passive or active) particles, inertia is typically negligibly small \cite{Zoettl:JPCM2016}.
An exception are effects due to hydrodynamic memory, the latter manifesting themselves in algebraically decaying long-time tails of correlation functions \cite{Felderhof:1983b, Franosch:2011, Huang:2011, Masters:1985, Lowe:1995};
however, effects due to such persistent memory are not included in the Langevin description of Brownian motion.
In the absence of an external torque, one sees from \cref{eq:euler-langevin} that the characteristic relaxation time of the angular velocity $\omega$ is given by $\tau_1 = I/\zeta_R$.
For a micron-sized sphere ($a=\SI{e-3}{mm}$) in water ($\eta/\rho = \SI{1}{mm^2/s}$), it holds $\tau_1 = \rho a^2/(15 \eta) \approx \SI{e-7}{s}$.
For times $t \gg \tau_1$, it is thus justified to neglect the inertial term $I \dot{\vec\omega}$ in \cref{eq:euler-langevin} relative to the dissipative term $\zeta_R\vec\omega$, yielding
\begin{equation}
  \zeta_R \, \vec \omega(t) =  \vec T_\ext(t) + \vec \xi(t) \,.
  \label{eq:euler-overdamped}
\end{equation}
This simple reasoning is possible since \cref{eq:euler-langevin} is of the Ornstein--Uhlenbeck type, in particular, since it is linear in $\vec \omega(t)$ and the noise $\vec \xi(t)$ enters additively.
The same result would be obtained from the formal solution for $\vec \omega(t)$, given in integral form \cite{Oksendal:Stochastic}, and taking the limit $\tau_1 \to 0$.

\Cref{eq:euler-overdamped} states that the angular velocity $\vec \omega(t)$ is statistically equivalent to a Gaussian white noise process with a deterministic bias $\vec T_\ext(t)$. More precisely, $\vec \omega(t)$ is a Gaussian process which satisfies
\begin{align}
  \langle \vec\omega(t) \rangle &= \zeta_R^{-1} \vec T_\ext(t) \,, \notag \\
  \langle \delta \vec\omega(t') \otimes \delta \vec\omega(t) \rangle &= 2 \DR \vec{1}\delta(t'-t) \,,
\label{eq:omega-moments}
\end{align}
upon introducing the unbiased part $\delta \vec\omega(t) = \vec\omega(t) - \zeta_R^{-1}\vec T_\ext(t)$
and the usual rotational diffusion constant $\DR = \kB T \zeta_R^{-1}$;
from here onward, $\langle \,\cdot\, \rangle$ refers to the noise average with respect to $\vec \omega(t)$.
We note that $\vec T_\ext(t)$ may depend on $\vec u(t)$, e.g., for a dipole-like alignment interaction between pairs of particles, which gives rise to the subtle question if (and in which sense) $\vec \omega(t)$ is independent of the orientation $\vec u(t)$.


It is now natural to interpret the kinematic equation \eqref{eq:kinematic} for the orientation $\vec u(t)$ as an SDE with a random angular velocity $\vec \omega(t)$.
We repeat the equation here:
\begin{equation}
  \dot{\vec u}(t) =  \vec\omega(t) \times \vec u(t)  \qquad \text{(Str)}\,,
  \label{eq:kinematic-Str}
\end{equation}
which is one form of the overdamped Langevin equation for rotational Brownian motion of a body-fixed axis $\vec u(t)$.
However, the white noise $\vec \omega (t)$ multiplies the variable $\vec u(t)$
and the re-interpretation of \cref{eq:kinematic} is yet ill-defined unless we specify the stochastic calculus
\footnote{We emphasize that we have left open this question when solving \cref{eq:euler-langevin}.},
i.e., unless we give a meaning to the time integral $\int_0^t \vec\omega(s) \times \vec u(s) \,\diff s$.
This is a modeling decision that needs problem-specific insight.
Recalling that $\vec u(t)$ is a unit vector, \cref{eq:kinematic-Str} describes a diffusion process on a constraining manifold, and the mathematically natural choice in this situation is to interpret the time integral as a (Fisk--)Stratonovich integral (see chap.~35 in ref.~\citenum{Kallenberg:Probability}).
After having fixed the stochastic integral, the physical soundness of the model must be verified, e.g., by testing detailed balance in equilibrium (see below, \cref{sec:detailed-balance}).

In the physics literature, the Stratonovich interpretation of \cref{eq:kinematic-Str} is often (tacitly and, in this case, correctly) assumed.
In the case of translational motion, rigorous arguments were established for the adiabatic elimination of linear momentum from underdamped Langevin equations \cite{Ryter:1981,Sancho:1982},
which can also be performed systematically using the the Fokker--Planck equation of the underdamped problem (Kramers equation) \cite{Risken:FokkerPlanck,Risken:1981}.
In principle, these procedures can be adapted to the present underdamped problem in $(\vec u, \vec \omega)$ [\cref{eq:kinematic,eq:euler-langevin}] to eliminate the fast variable $\vec \omega$.
Although both routes have recently been followed for two-dimensional rotational motion of ABPs \cite{Milster:EPJST2017},
they appear to be formidable tasks in three dimensions.

For the numerical integration of \cref{eq:kinematic-Str}, a direct application of the widespread Euler--Maruyama algorithm, which is a simple and robust algorithm for Itō SDEs, breaks the conservation of the norm $|\vec u(t)|$.
This can be fixed \emph{a posteriori} by rescaling $\vec u(t)$ after every integration step \cite{Tao:2005}\footnote{%
The algorithm due to Briels \emph{et al.} \cite{Tao:2005} is currently implemented in the molecular dynamics simulation software LAMMPS (since release from 29 September 2021), see the command ``fix Brownian/sphere'' at \url{https://docs.lammps.org/fix_brownian.html}.
}, but such an approach appears conceptually unsatisfying.
Instead, one should use a suitable discretization scheme for the Stratonovich integral, e.g., the stochastic Milstein algorithm; yet, the correct implementation of such schemes is more demanding \cite{KloedenPlaten:NumericalSolution,Waszkiewicz:SPPC2023}.
An alternative is to mix different stochastic calculi: one may cast the Stratonovich form \eqref{eq:kinematic-Str} into an equivalent form that invokes the Itō integral and which is thus suitable for the Euler--Maruyama scheme \cite{Risken:FokkerPlanck,Coffey:LangevinEquation,Oksendal:Stochastic};
this form includes a noise-induced term appearing as a drift [see below, \cref{eq:kinematic-Ito}].
We note that all of these integration schemes preserve the normalization of $\vec u(t)$ only asymptotically for a vanishing integration time step, $\Delta t \to 0$.

\subsection{Two-dimensional case.}

For motion in two dimensional space, which is widely considered in the theoretical literature on active particles,
the issue of the multiplicative noise can be circumvented since the angular velocity has only a single component.
Writing $\vec\omega = (0, 0, \omega)$, the component $u_z$ of $\vec u = (u_x, u_y, u_z)$ remains unchanged under the dynamics
so that we can set $u_z = 0$ and parametrise the remaining two components $(u_x, u_y) = (\cos(\phi), \sin(\phi))$ in terms of the angle $\phi \in [-\pi,\pi)$.
The evolution of $\phi(t)$ follows from \cref{eq:kinematic}, which reads for the first vector component:
$
  \dot u_x(t) = - \omega\, u_y(t) .
$
Combining with the ordinary chain rule,
$
  \dot u_x(t) = - \sin(\phi(t)) \, \dot \phi(t) ,
$
yields:
\begin{equation}
  \dot{\phi}(t) = \omega(t) \,.
  \label{eq:kinematic-2d}
\end{equation}
Passing to the overdamped limit, we replace $\omega(t)$ with the white noise given in \cref{eq:euler-overdamped},
which enters additively.
In fact, \cref{eq:kinematic-2d} is the simplest form of an SDE, which has an unambiguous solution with any stochastic calculus.
We conclude that $\phi(t)$ is a Brownian motion on the one-dimensional torus $[-\pi,\pi)$, i.e.,
$\vec u(t)$ performs a Brownian motion on the unit circle in the $xy$-plane.
The same result would be obtained when starting from \cref{eq:kinematic-Str} and observing that the classical (Leibniz) chain rule still applies for the Stratonovich calculus.

\subsection{Itō form of the overdamped equation}

For analytical and numerical work the use of Itō's calculus has certain advantages and, to this end, \cref{eq:kinematic-Str} needs to be cast into its Itō form, which reads:
\begin{equation}
  \dot{\vec u}(t) =  \vec\omega(t) \times \vec u(t) - \tauR^{-1} \vec u(t) \mathrlap{\qquad  \text{(It\=o)}}
  \label{eq:kinematic-Ito}
\end{equation}
with $\tauR^{-1} := (d-1)\DR$ for motion in $d$-dimensional space ($d=2,3$).
The transformation is straightforward \cite{Oksendal:Stochastic}, but a bit tedious due to the vector cross product on the r.h.s.\ of \cref{eq:kinematic-Str} and will be given below, 
after an alternative construction based on geometric considerations.
We will also show that, due to the specific rules of Itō calculus, the additional term
$- \tauR^{-1} \vec u(t)$ does not generate a drift, but is necessary to preserve the norm $|\vec u(t)|$ and to yield the orientational autocorrelation function correctly.
Furthermore, the solution $\vec u(t)$ is non-anticipating with respect to the noise $\vec \omega(t)$, i.e., the (Itō) increments of the noise, $\int_t^{t+\Delta t} \vec \omega(s) \diff s$ are independent of $\vec u(s)$ for any $t \geq s$ and $\Delta t > 0$.
(This central statement has no simple analogue for the Stratonovich integral.)
As a consequence, the Itō--Langevin equation \eqref{eq:kinematic-Ito} is amenable to the straightforward Euler--Maruyama scheme for numerical integration \cite{KloedenPlaten:NumericalSolution}.
Below, we propose a geometric integration scheme that satisfies the constraint $|\vec u(t)|=1$ exactly for arbitrary $\Delta t$.


\section{Geometric construction of the overdamped Langevin equation}

The discussion in \cref{sec:heuristic} is given from the angle of a bottom-up approach, starting from a Hamiltonian many-body system (Brownian particle plus solvent particles) and systematically integrating out fast degrees of freedom.
Here, we adopt a geometric perspective and follow a top-down approach by postulating properties of (idealised) Brownian motion on the unit sphere, similarly as it is established for Brownian motion in flat, Euclidean spaces.
We then show that this geometric construction [cf.\ \cref{eq:inf_rotations}] is equivalent to the accepted SDE for overdamped rotational Brownian motion, \cref{eq:kinematic-Str}.

\subsection{Geometric construction and derivation of the Itō form}
\label{sec:geometric-derivation}

In the following, we adopt a geometric perspective to obtain \cref{eq:kinematic-Ito}.
The Wiener process $\vec W(t)$, i.e., free and idealised Brownian motion, in a flat, Euclidean space can be constructed as a Lévy process with independent, stationary, and Gaussian distributed increments (or: displacements) $\vec W(t') - \vec W(t)$ for $t' > t$. Analogously,
Brownian motion of the orientation vector $\vec u(t)$ on the (three-dimensional) unit sphere is a Lévy process with independent, stationary, and isotropic increments such that $\vec u(t') \cdot \vec u(t)$ samples a certain distribution [cf.~\cref{eq:theta-dist}] determined by the diffusion equation on the sphere [cf.~\cref{eq:FPE}].
Here, the dot product enters via the usual metric on the unit sphere: the Euclidean distance is replaced by the length of the great circle arc between $\vec u(t')$ and $\vec u(t)$, which is $\arccos(\vec u(t') \cdot \vec u(t))$.
One concludes that the evolution of $\vec u(t)$ emerges from the repeated action of infinitesimal random rotations
and we write:
\begin{equation}
 \vec u(t + \diff t) = \e^{\vec \omega(t) \diff t \cdot \vec J} \vec u(t) \,,
 \label{eq:inf_rotations}
\end{equation}
where $|\vec \omega(t)| \diff t$ and $\vec \omega(t)/|\vec \omega(t)|$ are the angle and the axis of rotation at time $t$, respectively.
The symbol $\vec J = (J_1, \dots, J_3)$ denotes the $3\times 3$ matrices $(J_i)_{jk} = -\epsilon_{ijk}$, given in terms of the Levi--Civita symbol. These matrices form a basis of the Lie algebra $\mathfrak{so}(3)$
such that $\vec \omega \cdot \vec J = \omega_i J_i$ is the skew-symmetric matrix representation of the axial vector $\vec \omega$,
where summation of repeated indices is implied; written explicitly in components $\vec \omega = (\omega_x, \omega_y, \omega_z)$:
\begin{equation}
    \vec \omega \cdot \vec J = \begin{pmatrix}
      0    & -\omega_z  & \omega_y \\
    \omega_z & 0    & -\omega_x \\
    -\omega_y  & \omega_x & 0
    \end{pmatrix}
\end{equation}
In particular, it holds
\begin{equation}
  \vec \omega \times \vec v = (\vec \omega \cdot \vec J) \vec v = (\vec v \cdot \vec J)^\top \vec \omega
  \label{eq:wedge-product}
\end{equation}
for any vectors $\vec \omega$ and $\vec v$.
The ansatz \eqref{eq:inf_rotations} resembles the McKean--Gangolli injection \cite{Chirikjian:StochasticsLieGroups} of a stochastic process, here from the Lie algebra $\mathfrak{so}(3)$ into its Lie group $SO(3)$.

Expanding the exponential in \cref{eq:inf_rotations} up to the order of $[\vec \omega(t)\diff t]^2$ yields, according to the rules of Itō calculus,
\begin{align}
  \vec u(t + \diff t) &= \left[\mathds{1} + \vec \omega(t)\diff t \cdot \vec J  + \tfrac{1}{2}(\vec \omega(t) \diff t \cdot \vec J)^2 \right] \vec u(t) \notag \\
    &= \vec u(t) + \vec \omega(t)\diff t \times \vec u(t) - 2 \DR \vec u(t) \diff t  \,,
  \label{eq:inf_rotations_expanded}
\end{align}
where we have used that
\begin{multline}
 (\vec \omega \diff t \cdot \vec J)^2 = \sum_{ij} \omega_i \omega_j \diff t^2 J_i J_j \\
 = \sum_{i} 2 \DR \diff t (J_i)^2 = -4 \DR \mathds{1} \diff t \,.
\end{multline}
\Cref{eq:inf_rotations_expanded} is the same as \cref{eq:kinematic-Ito}, which is seen by subtracting $\vec u(t)$ from both sides of \cref{eq:inf_rotations_expanded} and dividing it by the differential $\diff t$.

\subsection{Discussion of the drift term}
\label{sec:drift-term}

Before, we discuss the term $-\tauR^{-1}\vec u(t)$ on the r.h.s.\ of \cref{eq:kinematic-Ito}, which appears to describe a drift.
However, the term is normal to the constraining manifold and hence cannot generate a drift.
Instead, it is needed to satisfy the constraint, i.e., to preserve the normalization $|\vec u(t)| = 1$ under the dynamics of \cref{eq:kinematic-Ito}.
In order to prove this fact, we switch to the notation of stochastic (It\=o) differentials and write
$\diff \vec u(t) = \dot{\vec u}(t) \diff t$;
the expression $\vec \omega(t) \diff t$ is the differential of a scaled (and shifted
\footnote{The argument given holds also in the presence of $\vec T_\ext(t)$ in \cref{eq:omega-moments}.})
Wiener process.
Then, \cref{eq:omega-moments} implies $\omega_i(t) \diff t \, \omega_j(t) \diff t = 2 \DR \delta_{ij} \diff t$ for the Cartesian components $i$ and $j$ of $\vec \omega(t)$.
Employing Itō's formula, substituting $\diff \vec u(t)$ by means of \cref{eq:kinematic-Ito}, and omitting the time arguments for brevity, one finds:
\begin{align}
 \diff (\vec u \cdot \vec u) &= 2 \vec u \cdot \diff \vec u + \diff \vec u \cdot \diff \vec u \notag \\
  &\stackrel{\mathclap{\eqref{eq:kinematic-Ito}}}{=} - 2\tauR^{-1}  \vec u \cdot \vec u \diff t
      + (\vec\omega \diff t \times \vec u) \cdot (\vec\omega\diff t \times \vec u) \notag \\
  &= - 2\tauR^{-1}  |\vec u|^2 \diff t
      + \sum_i (\omega_i \diff t)^2 |\vec u|^2 \notag \\
      & \qquad -  \sum_{ij} \omega_i\diff t \, \omega_j\diff t  \, u_i \, u_j \notag \\
  &= - 2\tauR^{-1}  |\vec u|^2 \diff t
      + 6 \DR \diff t |\vec u|^2
      -  2\DR \diff t \, |\vec u|^2 \notag \\
  &= 2\left(2\DR - \tauR^{-1}\right) |\vec u|^2 \diff t \,.
\end{align}
%
%
Therefore, provided that $\tauR^{-1} = 2\DR$, it holds $\diff |\vec u(t)|^2 / \diff t = 0$ or, equivalently, $|\vec u(t)| = \mathrm{const}$.

In \cref{sec:FP-equation}, we derive the Fokker--Planck equation corresponding to \cref{eq:kinematic-Ito}.
From a different angle, it also proves that the term $-\tauR^{-1}\vec u(t)$ does not contribute to the probability flux.
(In fact, the \emph{absence} of this term would entail a spurious drift.)
Moreover, for $\vec T_\ext=0$, the equilibrium distribution of the orientation $\vec u$ is obtained to be uniform on the unit sphere, as required for free rotational diffusion.

\subsection{Orientational autocorrelation function}

The importance of the Itō term in \cref{eq:kinematic-Ito} becomes clear again when computing the autocorrelation function of the orientation, $C_1(t) := \langle \vec u(t) \cdot \vec u(0) \rangle$, which is a simple exercise in the absence of an external torque $\vec T_\ext(t)=0$.
Multiplication of \cref{eq:kinematic-Ito} by $\vec u(0)$, integrating over time, and averaging yields
\begin{equation}
  C_1(t) = \biggl\langle \int_0^t [\vec\omega(s) \times \vec u(s)] \cdot \vec u(0) \,\diff s \biggr\rangle
    - \tauR^{-1} \int_0^t C(s) \, \diff s \,.
\end{equation}
Recalling that $\vec \omega(s)$ is a white noise process [\cref{eq:omega-moments}] and independent of $\vec u(s)$ and $\vec u(0)$, the first term on the r.h.s.\ contains a properly formed Itō integral, which is zero on average.
(The Stratonovich integral does not share this property.)
Here, we have used that $\langle \omega(t) \rangle = \vec T_\ext(t)=0$.
The remaining integral equation has the expected solution
\begin{equation}
  C_1(t) = \exp(-t/\tauR) \,,
\end{equation}
which is seen, for example, by differentiating with respect to $t$ and solving the obtained ordinary differential equation for $C_1(t)$ with $C_1(0)=|\vec u(0)|^2 = 1$.

\subsection{Equivalence of the Itō and Stratonovich forms}
\label{sec:equivalence}

It remains to show the mathematical equivalence of \cref{eq:kinematic-Str,eq:kinematic-Ito}.
Generally, given a diffusion process $\vec X(t)$ in $\mathbb{R}^d$, which is driven by a standard Wiener process $\vec W(t)$ in $\mathbb{R}^n$
scaled by an $d \times n$ matrix-valued coefficient function $\sigma(\vec X(t))$,
the corresponding Stratonovich integral, denoted by $\hbox{}\circ \diff \vec W(t)$ as usual, is related to the Itō integral via \cite{Kallenberg:Probability,Gardiner:Stochastic}
\begin{multline}
 \int_0^T \sigma_{ij}(\vec X(t)) \circ \diff W_j(t)
  = \int_0^T \sigma_{ij}(\vec X(t)) \, \diff W_j(t) \\
     + \frac{1}{2} \int_0^T [\partial_k \sigma(\vec X(t))]_{ij} \sigma_{kj}(\vec X(t)) \, \diff t \,,
 \label{eq:Str-integral}
\end{multline}
for some time $T > 0$ and $\partial_k$ denoting the partial derivative w.r.t.\ to the component $X_k$;
the summation convention is applied to repeated indices.
Omitting the integral signs in \cref{eq:Str-integral} (formally, taking a derivative w.r.t.\ $T$)
yields the differential form of this relation in the common short-hand notation:
\begin{equation}
 \sigma_{ij}(\vec X) \circ \diff W_j =
  \sigma_{ij}(\vec X) \, \diff W_j + \frac{1}{2} [\partial_k \sigma(\vec X)]_{ij} \sigma_{kj}(\vec X) \, \diff t \,.
 \label{eq:Str-differential}
\end{equation}
In order to apply it to \cref{eq:kinematic-Str}, one identifies $\vec X(t)$ with $\vec u(t)$ and $\diff \vec W(t)$ with $(2\DR)^{-1/2} \delta \vec\omega(t) \,\diff t$.
The noise strength $\sigma(\cdot)$ is $3\times 3$ matrix-valued:
$\sigma(\vec u) = -(2\DR)^{1/2} \, \vec u \cdot \vec J$,
which has the derivatives
$\partial_k \sigma(\vec u) = - (2\DR)^{1/2} J_k$.
This is seen by rewriting the cross product in matrix form [\cref{eq:wedge-product}],
$(\delta \vec \omega \times \vec u) \, \diff t = -(\vec u \cdot \vec J) \, \delta \vec \omega \, \diff t$.
Therefore,
\begin{align}
 [-\vec u(t) \cdot \vec J]_{ij} \circ \delta \omega_j(t) \, \diff t \hspace{-5em} \notag \\
 &= [-\vec u(t) \cdot \vec J]_{ij} \, \delta \omega_j(t) \, \diff t \notag \\
    & \qquad + \frac{2\DR}{2} (-J_k)_{ij} [-\vec u(t) \cdot \vec J]_{kj} \, \diff t  \notag \\
 &= \delta \vec \omega(t) \, \diff t \times \vec u(t)
  - 2 \DR \vec u(t) \, \diff t \,.
  \label{eq:Str-vs-Ito}
\end{align}
For the last step, we have simplified as follows:
\begin{multline}
 (-J_k)_{ij} [-\vec u(t) \cdot \vec J]_{kj}
 = (J_k)_{ij} (J_l)_{kj} u_l(t) \\
 = \epsilon_{kij} \epsilon_{lkj} u_l(t)
 = (\delta_{kl} \delta_{ik} - \delta_{kk} \delta_{il}) u_l(t)
 = -2 u_i(t) \,;
  \label{eq:J_u_J}
\end{multline}
$\delta_{ij}$ is the Kronecker symbol.
After division of \cref{eq:Str-vs-Ito} by $\diff t$, it is clear that, for $\tauR^{-1} = 2\DR$, the r.h.s.\ of the Stratonovich SDE \eqref{eq:kinematic-Str} corresponds to the r.h.s.\ of the Itō SDE \eqref{eq:kinematic-Ito} as claimed.
We note that the external torque contributes to the systematic drift and does not enter $\sigma(\vec u)$ and, thus, not the transformation \eqref{eq:Str-differential} of the stochastic integral; in particular, $\delta\vec \omega(t)$ in \cref{eq:Str-vs-Ito} may be replaced by $\vec \omega(t)$.

\subsection{Two-dimensional case}

The geometric construction of \cref{eq:kinematic-Ito} has a straightforward generalization to $d=2$ dimensions.
We note that the angular velocity $\vec \omega(t)$, being an axial vector, has $n := d(d-1)/2$ components since it can be represented as a skew-symmetric $d\times d$ matrix.
In particular for $d=2$, it has only a single component $\omega_z$ (which may be chosen as the $z$-component in a three-dimensional embedding of the motion).
Correspondingly, the Lie algebra $\mathfrak{so}(2)$ is generated by a single matrix, e.g.,
$J_z = \begin{pmatrix} 0 & -1 \\ 1 & 0 \end{pmatrix}$.
With this, the cross product in \cref{eq:wedge-product} is to be understood as
\begin{equation}
  \vec \omega \times \vec v = (\vec \omega \cdot \vec J) \vec v = \omega_z J_z \vec v
\end{equation}
for two-dimensional vectors $\vec v$.
The reasoning from \crefrange{eq:inf_rotations}{eq:inf_rotations_expanded} remains unchanged except for the last step,
where one has to replace $-\sum_i (J_i)^2 = 2 \cdot \mathds{1}$ by $-(J_z)^2 = \mathds{1}$.
Hence, \cref{eq:kinematic-Ito} applies also in $d=2$ dimensions, but with $\tauR^{-1} = \DR$.

For the proof of the equivalence of the Itō and Stratonovich forms (\cref{sec:equivalence}), the following changes apply:
The noise strength $\sigma(\vec u) = -(2\DR)^{1/2} J_z \vec u$ is a $2\times 1$ matrix
with derivative $\partial_k \sigma(\vec u) = - (2\DR)^{1/2} J_z \vec e_k$; here, $\vec e_k$ are the Cartesian unit vectors,
$(\vec e_k)_j = \delta_{kj}$.
Second, the calculation in \cref{eq:J_u_J} is replaced by
\begin{multline}
 (2\DR)^{-1} [\partial_k\sigma(\vec u)]_{ij} \sigma(\vec u)_{kj}
  = (-J_z \vec e_k)_{ij} [-J_z \vec u(t)]_{kj} \\
  = (J_z)_{ik} (J_z)_{kl} u_l(t)
  = [(J_z)^2 \vec u(t)]_i = - u_i(t) \,,
\end{multline}
where the sums run over rows $k,l \in \{1,2\}$ and column $j=1$.

\section{Geometric numerical integration scheme}
\label{sec:geometric}

\subsection{Algorithm}
\label{sec:algorithm}

Above, we have shown that \cref{eq:inf_rotations} is equivalent to \cref{eq:kinematic-Ito}.
The latter is a standard Itō SDE and may be integrated numerically with the Euler--Maruyama scheme or some higher-order scheme \cite{KloedenPlaten:NumericalSolution,Gardiner:Stochastic},
introducing some finite integration time step $\Delta t$.
However, these schemes would satisfy the constraint $|\vec u(t)|=1$ only asymptotically, for $\Delta t \to 0$.

Alternatively, \cref{eq:inf_rotations} suggests to implement a geometric integration scheme for Brownian motion on the unit sphere as a sequence of rotations with finite random angles, which reads for a single integration step:
\begin{equation}
 \vec u(t + \Delta t) = \e^{\Delta \vec \Omega(t) \cdot \vec J} \vec u(t)
 \label{eq:finite_rotation}
\end{equation}
for yet to be determined random vectors $\Delta \vec \Omega(t)$,
recalling that $\e^{\theta \unitvec n \cdot \vec J}$ is a rotation matrix with axis $\unitvec n := \Delta \vec \Omega(t) / |\Delta \vec \Omega(t)|$ and angle $\theta := |\Delta \vec \Omega(t)|$.
The explicit action of this rotation on the unit vector $\vec u=\vec u(t)$ is given for $d=3$ by Rodrigues formula,
\begin{multline}
  \vec u(t + \Delta t) = \cos(\theta) \vec u
    - \sin(\theta) \, \vec u \times \unitvec n \\
    + [1 - \cos(\theta)] (\vec u \cdot \unitvec n) \unitvec n \,.
  \label{eq:Rodrigues}
\end{multline}
Nowadays, the trigonometric functions are evaluated by dedicated special function units
on typical computing hardware, either CPUs or GPUs, and the appearance of such functions in the algorithm is not a performance issue anymore.

It remains to specify the statistics of $\Delta \vec \Omega(t)$.
Comparison of \cref{eq:finite_rotation} and \cref{eq:inf_rotations} yields for small time steps that
\begin{equation}
  \Delta \vec \Omega(t) \simeq \vec \omega(t) \Delta t \,; \quad \Delta t \to 0 .
  \label{eq:Omega_asymptotics}
\end{equation}
With this choice, time integration of \cref{eq:omega-moments} suggests that
$\Delta \vec \Omega(t)$ is a Gaussian vector with independent components.
Clearly, the rotation of $\vec u$ about its own axis has no effect, which also follows from setting $\unitvec n = \vec u$ in \cref{eq:Rodrigues}. Thus, it is sufficient to choose $\Delta \vec\Omega$ in the plane perpendicular to $\vec u$:
\begin{equation}
  \Delta \vec\Omega = \Omega_1 \vec e_1 + \Omega_2 \vec e_2
  \label{eq:Omega-components}
\end{equation}
for unit vectors $(\vec e_1, \vec e_2, \vec u)$ forming a trihedron.
The random coefficients $\Omega_1$ and $\Omega_2$ are independent Gaussian variables with means and variances given by
\begin{equation}
  \langle \Omega_i \rangle  = \vec e_i \cdot \zeta_R^{-1} \vec T_\ext(t)
  \quad \text{and} \quad
  \Var[\Omega_i] = 2 \DR \Delta t
  \label{eq:Omega-moments}
\end{equation}
for $i=1,2$, also including a possible external torque.

\begin{table}
\tabcolsep=1em
\begin{tabularx}{.9\linewidth}{lX}
\toprule
Inputs: & $\vec u = \vec u(t)$, $\vec T_\ext(t)$, $\DR = \kB T \zeta_R^{-1}$, $\Delta t$ \\
Output: & $\vec u(t + \Delta t)$ \\
\midrule
\multicolumn{2}{c}{\parbox{.8\linewidth}{%
\begin{enumerate}[label={(\roman*)}]
 \item construct the vector $\unitvec e_1 \perp \vec u$ and let $\unitvec e_2 = \vec u \times \unitvec e_1$
 \item draw normally distributed random coefficients $\Omega_1$ and $\Omega_2$ such that
  $
    \langle \Omega_i \rangle  = \vec e_i \cdot \zeta_R^{-1} \vec T_\ext(t)
  $
  and
  $\Var[\Omega_i] = 2 \DR \Delta t$

 \item compute $\unitvec n = \Delta \vec \Omega / |\Delta\vec\Omega|$ and $\theta = |\Delta\vec\Omega|$
  and evaluate $\sin(\theta)$ and $\cos(\theta)$
 \item obtain
  $
    \vec u(t + \Delta t) = \cos(\theta) \vec u - \sin(\theta) \, \vec u \times \unitvec n
  $
\end{enumerate}
}} \\
\bottomrule
\end{tabularx}
\caption{Algorithm for the geometric integration scheme [\cref{eq:finite_rotation}] using Gaussian distributed random rotations [\cref{eq:Omega-moments}].}
\label{tab:algorithm}
\end{table}

In summary, each time step of the geometric integration scheme [\cref{eq:finite_rotation,eq:Rodrigues,eq:Omega_asymptotics,eq:Omega-moments}] consists of the algorithm listed in \cref{tab:algorithm}.
In step (iv), we have simplified \cref{eq:Rodrigues}, noting that $\vec u \perp \unitvec n$.
Optionally, after step (iv), one may normalize the unit vector $\vec u(t +\Delta t)$ to avoid a possible drift of $|\vec u|$ after a large number of integration steps, which can result from round-off errors due to limited floating-point precision. We emphasize again that the time discretization in this scheme preserves the normalization exactly.

\subsection{Estimation of the numerical error}

The form of \cref{eq:finite_rotation} admits an exact representation of the solution of the SDE \eqref{eq:inf_rotations}
in the sense that any given pair of $\vec u(t)$ and $\vec u(t+\Delta t)$ defines a rotation vector $\Delta \vec\Omega(t)$.
However,
\begin{equation}
  \Delta \vec \Omega(t) \neq \int_t^{t+\Delta t} \vec \omega(s) \diff s
\end{equation}
in general and, thus,
\cref{eq:Omega_asymptotics} describes merely an approximation for finite $\Delta t$;
the reason behind being the non-commutativity of finite rotations:
\begin{equation}
  \e^{\Delta \vec \Omega_A \cdot \vec J} \e^{\Delta \vec \Omega_B \cdot \vec J} \neq \e^{(\Delta \vec \Omega_A + \Delta \vec \Omega_B) \cdot \vec J}
\end{equation}
for $\Delta \vec \Omega_A$ and $\Delta \vec \Omega_B$ of subsequent integration intervals.

In order to estimate the discretization error, we compare the numerically obtained distribution of $\vec u(t+\Delta t)$ with the propagator of the dynamics of $\vec u(t)$, putting $\vec T_\ext(t)=0$ for simplicity.
The free propagator is known exactly from the solution of the corresponding Fokker--Planck equation [cf.\ \cref{sec:FP-equation}] and, assuming that $\vec u(0)$ is sampled in equilibrium, it reads in terms of the correlation functions \cite{BernePecora:DynamicLightScattering}
\begin{align}
 C_\ell(t) &:= \langle P_\ell\boldsymbol(\vec u(t) \cdot \vec u(0)\boldsymbol) \rangle \notag \\
  &= \e^{-\ell(\ell+1) \DR \Delta t} \,, \qquad \ell=0,1,2,\dots \,,
  \label{eq:corr-exact}
\end{align}
where $P_\ell(\cdot)$ denotes the Legendre polynomial of degree~$\ell$.
Since $\vec u(t)$ is a Markov process, it is sufficient to consider the first step from $t=0$ to $t=\Delta t$.
The numerical scheme [\cref{eq:Rodrigues}] yields $\vec u(t) \cdot \vec u(0) = \cos(\theta)$ and thus
for the ``numerical'' propagator:
\begin{equation}
  \what C_\ell(t) = \langle P_\ell\boldsymbol(\cos(|\Delta \vec \Omega|)\boldsymbol) \rangle_{\Delta \vec \Omega} \,,
  \label{eq:corr-numerical}
\end{equation}
where we have used that $\vec n \perp \vec u(0)$ by construction and $|\vec u(0)|=1$;
the average is taken with respect to the distribution of $\Delta \vec \Omega$ given in \cref{eq:Omega-moments}.
Inserting the Gaussian distributions for $\Omega_1$ and $\Omega_2$ and transforming to the (unwrapped) angle $\theta = (\Omega_1^2 + \Omega_2^2)^{1/2}$, one obtains an explicit expression for the propagator corresponding to \cref{eq:Rodrigues}:
\begin{equation}
  \what C_\ell(t) = \int_0^\infty P_\ell\boldsymbol(\cos(\theta)\boldsymbol) \, \frac{1}{2 \DR t} \,\e^{-\theta^2 / 4 \DR t} \, \theta \diff \theta  \,.
  \label{eq:corr-numerical-gaussian}
\end{equation}
This integral can be written in terms of Dawson's $F$-function \cite{Abramowitz:MathFunc}. For our purposes, however, it suffices to expand the integrand in $\theta=0$ since the distribution is sharply peaked for small $\DR t \ll 1$:
\begin{align}
  \what C_\ell(t)
  &= \int_0^\infty \left[1 - \frac{\ell(\ell+1)}{4}\theta^2 + \frac{1}{24} a_2(\ell) \theta^4 + O(\theta^6)\right] \, \notag \\
  & \qquad \times \frac{1}{4 \DR t} \, \e^{-\theta^2 / 4 \DR t} \, \diff(\theta^2) \notag \\[.5ex]
  &= 1 - \ell (\ell+1) \DR t + \frac{4}{3} a_2(\ell) (\DR t)^2 + O\mleft((\DR t)^4\mright)
\end{align}
where $\alpha_2(\ell) = \ell (\ell+1) [3 \ell (\ell+1)-2] / 8$. From the comparison with \cref{eq:corr-exact} and writing $\Delta t$ again instead of $t$, one finds the relative numerical error of the correlation functions $C_\ell(t)$ after one integration step of length $\Delta t$:
\begin{equation}
 \frac{\what C_\ell(\Delta t) - C_\ell(\Delta t)}{C_\ell(\Delta t)}
    = - \frac{\ell (\ell+1)}{3} \, (\DR \Delta t)^2 + O\mleft((\DR \Delta t)^3\mright) \,,
  \label{eq:corr-error-single}
\end{equation}
asymptotically as $\DR \Delta t \to 0$.
In order to integrate $\vec u(0)$ up to time $t= N \Delta t$ one performs the propagation step in \cref{eq:finite_rotation} successively $N$ times.
Correspondingly, the relative discretization error [\cref{eq:corr-error-single}] adds up $N$ times so that the
overall error obeys
\begin{equation}
 |\what C_\ell(t) - C_\ell(t)| \simeq \frac{\DR \Delta t}{3} \,  \ell (\ell+1) \DR t \, \e^{-\ell(\ell+1) \DR t}
 \label{eq:corr-error}
\end{equation}
upon $\DR \Delta t \to 0$, which we have verified in numerical tests.
Moreover, the $t$-dependent factor is bounded and we can estimate the error for arbitrary times:
\begin{equation}
 \sup_{t \in [0,\infty)} |\what C_\ell(t) - C_\ell(t)| = \frac{\DR \Delta t}{3 \e} + O\mleft((\DR \Delta t)^2\mright) \,.
\end{equation}
One concludes that the geometric integrator shows a globally weak convergence of order 1 in the integration time step $\Delta t$.

\subsection{A weakly exact integration scheme}

A weakly exact integration scheme (i.e., one that yields the correct statistics of the solution \cite{Gardiner:Stochastic,KloedenPlaten:NumericalSolution}) is obtained upon replacing \cref{eq:Omega-moments} by the true statistics of $\Delta \vec \Omega$, which tends to a Gaussian only for $\DR \Delta t \ll 1$.
In particular, a series representation of the distribution of the rotation angle $\theta = |\Delta \vec \Omega|$, here restricted to $0\leq \theta \leq \pi$, follows from its angular moments [\cref{eq:corr-exact}]:
\begin{equation}
  p(\theta,t) = \sum_{\ell \geq 0} \left(\ell + \tfrac{1}{2}\right) C_\ell(t) P_\ell\boldsymbol(\cos(\theta)\boldsymbol) \sin(\theta) \,.
  \label{eq:theta-dist}
\end{equation}
With this, $C_\ell(t) = \int_0^\pi P_\ell\boldsymbol(\cos(\theta)\boldsymbol) \, p(\theta,t) \, \diff \theta$
and reweighting the angle $\theta$ in step (iii) of the algorithm in \cref{tab:algorithm} such that it samples $p(\theta)$ yields numerical correlation functions $\what C_\ell(t)$ [\cref{eq:corr-numerical}] which are equal to the exact solution $C_\ell(t)$ for all $t$.
A truncation of the series for $p(\theta,t)$ by keeping only terms with $\ell \leq \ell_\text{max}$ yields
$\what C_\ell(t) = C_\ell(t)$ for $\ell \leq \ell_\text{max}$ and
$\what C_\ell(t) = 0$ otherwise.
We note that $p(\theta,t)$ is similar in shape, but different from the van-Mises-Fisher distribution,
$p_\text{vMF}(\theta,t) \propto \exp(\cos(\theta)/2\DR t) \sin(\theta)$,
which decays too slowly as $\theta \to \pi$ and would yield a larger integration error than given in \cref{eq:corr-error-single}.

Numerically, the direct sampling from the (truncated or untruncated) distribution \eqref{eq:theta-dist}
may be implemented via inverse transform sampling:
$F^{-1}(Z)$ samples the density $p(\theta,\Delta t)$ if $Z$ is a uniformly distributed random variable on the interval $[0,1]$ and the function $F^{-1}(\cdot)$ is the (numerically pre-computed) inverse of the cumulative distribution function of $\theta$, i.e., $F(\theta) = \int_0^\theta p(\theta',\Delta t) \, \diff\theta'$ for $\theta \in [0,\pi]$.
Alternatively, one can imagine to start from a Gaussian distributed variable $Z$ and to use $Z=|\Delta \vec \Omega|$ from the original step (iii) of the algorithm.
An efficient implementation of this exact variant of the integration scheme is left for future research.

\subsection{Comparison to Briels' projection scheme}

\begin{figure}
  \includegraphics{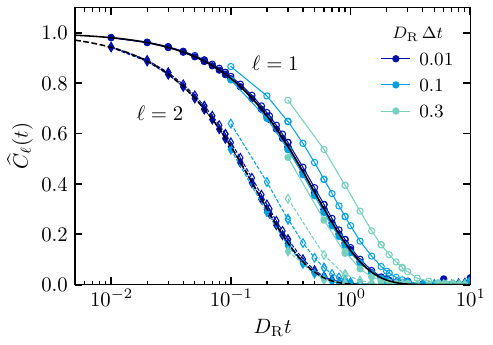}
  \caption{Numerically obtained correlation functions $\what C_\ell(t)$ for $\ell=1$ (circles) and $\ell=2$ (diamonds)
  using the geometric integrator (algorithm in \cref{tab:algorithm}, closed symbols) and the projection scheme [Eqs.~\eqref{eq:adhoc-scheme}, open symbols], which combines Euler--Maruyama discretization of the Langevin equation \eqref{eq:kinematic-Str} with \emph{a posteriori} normalization of the orientation vector.
  Colors distinguish different integration time steps $\Delta t$, numerical data are available only for correlation times $t$ that are a multiple of $\Delta t$, and colored lines serve as guides to the eye.
  The solid and dashed black lines show the exact solutions $C_\ell(t)$ for $\ell = 1$ and $\ell = 2$, respectively [\cref{eq:corr-exact}].
  }
  \label{fig:correlations}
\end{figure}

For comparison, we also give a brief analysis of the numerical scheme suggested by Briels \emph{et al.} \cite{Tao:2005}, which combines an Euler--Maruyama step for the Stratonovich SDE \eqref{eq:kinematic-Str} with a post-normalization step of the unit vector;
the second step corresponds to an orthogonal projection onto the constraining manifold, $|\vec u(t+\Delta t)| = 1$.
This scheme corresponds to the geometric algorithm in \cref{tab:algorithm} with step (iv) replaced by:
\begin{subequations}
\label{eq:adhoc-scheme}
\begin{align}
  \vec u' &= \vec u(t) + \Delta \vec \Omega(t) \times \vec u(t) \,,
  \label{eq:adhoc-scheme-a} \\
  \vec u(t + \Delta t) &= \vec u' / |\vec u'| \,;
  \label{eq:adhoc-scheme-b}
\end{align}
\end{subequations}
as before, $\Delta \vec \Omega(t)$ is a Gaussian random vector satisfying \cref{eq:Omega-moments}.
The first equation may also be obtained from \cref{eq:Rodrigues} by expanding in $\theta=0$ and keeping only terms $O(\theta)$.

A calculation similar to the one in \cref{sec:drift-term} shows that the post-normalization in \cref{eq:adhoc-scheme-b} is equivalent (in distribution) to adding the Itō drift term to \cref{eq:adhoc-scheme-a} up to linear order in $\Delta t \to 0$:
\begin{align}
  \vec u(t + \Delta t) &= \frac{\vec u(t) + \Delta \vec \Omega(t) \times \vec u(t)}{(1 + |\Delta \vec \Omega(t) \times \vec u(t)|^2)^{1/2}} \notag \\
  & \simeq \vec u(t) + \Delta \vec \Omega(t) \times \vec u(t) - 2 \DR \vec u(t) \Delta t \,.
  \label{eq:adhoc-linear}
\end{align}
We conclude that the projection scheme has the same convergence properties as the Euler--Maruyama scheme for the Itō--Langevin equation [\cref{eq:kinematic-Ito}]; in particular, it also exhibits weak convergence of order~1.
However, the magnitude of the error is significantly larger compared to the geometric scheme (\cref{tab:algorithm}). This is demonstrated by numerical results for the correlation functions $\what C_1(t)$ and $\what C_2(t)$ obtained from both schemes (\cref{fig:correlations}).
For an integration time step of $\Delta t = 0.1 \DR^{-1}$, the results from the projection scheme deviate already considerably from the exact solutions, whereas merely insignificant deviations are visible for $\Delta t = 0.3 \DR^{-1}$ using the geometric scheme.

\begin{figure}
  \includegraphics{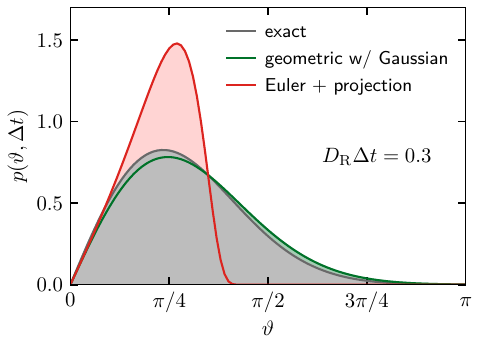}
  \caption{Probability distribution $p(\theta, \Delta t)$ of the angle $\theta$ between the initial and final orientations after an integration time step of length $\Delta t=0.3 \DR^{-1}$ for the geometric scheme with Gaussian rotational vectors [\cref{eq:theta-dist-gauss}, green line] and for Briels' projection scheme, combining an Euler--Maruyama step with an orthogonal projection [\cref{eq:theta-dist-adhoc}, red line].
  The gray line shows the exact propagator of rotational Brownian motion [\cref{eq:theta-dist}].
  }
  \label{fig:propagators}
\end{figure}

For the geometric scheme with Gaussian rotation vectors, the wrapped rotation angle $0 \leq \theta \leq \pi$ is distributed with  probability density [see \cref{eq:corr-numerical-gaussian}]
\begin{align}
  p_\text{Gauss}(\theta,\Delta t)
  &= \int_0^\infty \frac{\theta'}{2 \DR \Delta t} \,\e^{-(\theta')^2 / 4 \DR \Delta t} \notag \\
  & \qquad \times \delta\boldsymbol(\cos(\theta) - \cos(\theta')\boldsymbol) \,|\sin(\theta')| \diff \theta'\,. \notag \\
  &= \sum_{k =0}^\infty \frac{\theta + 2\pi k}{2 \DR \Delta t} \,\e^{-(\theta + 2\pi k)^2 / 4 \DR \Delta t} \notag \\
  & \qquad + \sum_{k =1}^\infty \frac{2\pi k-\theta}{2 \DR \Delta t} \,\e^{-(2\pi k - \theta)^2 / 4 \DR \Delta t} \,.
  \label{eq:theta-dist-gauss}
\end{align}
For most practical purposes, it is sufficient to restrict the sum to the $k=0$ term;
the next-to-leading order term ($k=1$) is suppressed by a factor of $\exp(-\pi^2/4\DR \Delta t)$.
%

In the case of the projection scheme, the final rotation angle is given via [see \cref{eq:adhoc-linear}]:
\begin{align}
  \cos(\theta) &= \vec u(t + \Delta t) \cdot \vec u(t) \notag \\
  &= \bigl(1 + |\Delta \vec \Omega(t)|^2\bigr)^{-1/2} \,,
\end{align}
making use of $|\vec u(t)|=1$ and $\Delta\vec \Omega(t) \perp \vec u(t)$.
One finds for the distribution of $z=\cos(\theta)$:
\begin{align}
 \bar p_\text{proj}(z,\Delta t) &= \int_0^\infty \delta\mleft(z - \frac{1}{\sqrt{1+\theta^2}}\mright) \,
  \frac{\e^{-\theta^2 / 4 \DR \Delta t}}{2 \DR \Delta t} \, \theta \diff \theta \notag \\
  &= \frac{z^{-3}}{2\DR \Delta t} \e^{-(z^{-2} - 1)/4 \DR \Delta t} \,, \quad z > 0.
\end{align}
Zero or negative values, $z \leq 0$, cannot be attained in this scheme since $\vec u'$ is in the tangent plane of the constraint at point $\vec u$ and thus its projection cannot reach the lower half-sphere.
It follows that the probability density of the angle $0 \leq \theta < \pi/2$ reads:
\begin{equation}
  p_\text{proj}(\theta, \Delta t) = 
  \frac{\sin(\theta)}{2\DR \Delta t \cos(\theta)^3} \,\exp\mleft(-\frac{\cos(\theta)^{-2} - 1}{4 \DR \Delta t}\mright) \,.
  \label{eq:theta-dist-adhoc}
\end{equation}
One observes that, for small rotation angles,
\begin{equation}
  p_\text{proj}(\theta \to 0, \Delta t) = p_\text{Gauss}(\theta, \Delta t)
    \, \bigl[1 + O(\theta^2 / \DR \Delta t)\bigr] \,.
\end{equation}

The obtained distributions $p(\theta,\Delta t)$ are the propagators of the various integration schemes investigated.
Their graphical comparison in \cref{fig:propagators} for $\Delta t = 0.3 \DR^{-1}$
shows that the geometric scheme with Gaussian rotation vectors (\cref{tab:algorithm}) closely follows the exact distribution [\cref{eq:theta-dist}], despite the comparably large time step.
In contrast, the propagator of the projection scheme deviates significantly, which underpins the observations made before for the correlation functions (\cref{fig:correlations}).

\section{Extension to axisymmetric particles}
\label{sec:axisymmetric}

For the extension to axisymmetric particles, let $\vec u(t)$ denote the symmetry axis under rotations. Then, the friction tensor and moment of inertia in the spaced-fixed frame attain the forms \cite{HappelBrenner:Hydrodynamics}:
\begin{align}
  \tens \zeta_R(t) &= \zeta_R^\parallel \mathcal{P}_{\vec u(t)} + \zeta_R^\perp \mathcal{Q}_{\vec u(t)} \,, \\
  \tens I(t) &= I_\parallel \mathcal{P}_{\vec u(t)} + I_\perp \mathcal{Q}_{\vec u(t)} \,,
\end{align}
where $\mathcal{P}_{\vec u} = \vec u \otimes \vec u$ is the orthogonal projector on the given direction $\vec u$
and $\mathcal{Q}_{\vec u} = \mathds{1} - \mathcal{P}_{\vec u}$ its complement.
For Brownian particles, the argument of time scale separation leading to \cref{eq:euler-langevin} applies analogously since the matrix norms of $\tens \zeta_R(t)$ and $\tens I(t)$ remain unchanged under the rotational transformations.
Thus, in the overdamped limit, \cref{eq:euler-overdamped} holds with a time-dependent, stochastic friction tensor $\tens\zeta_R(t)$.
Multiplying the equation with the reciprocal tensor,
\begin{equation}
  \tens \zeta_R(t)^{-1} = (\zeta_R^\parallel)^{-1} \mathcal{P}_{\vec u(t)}
  + (\zeta_R^\perp)^{-1} \mathcal{Q}_{\vec u(t)} \,, \\
\end{equation}
yields
\begin{equation}
  \vec \omega(t) =  \tens \zeta_R(t)^{-1} \vec T_\ext(\vec u(t), t)
    + \tens \zeta_R(t)^{-1} \vec \xi(t) \,.
\end{equation}
Here, we have also included a possible dependence of the external torque on the instantaneous orientation $\vec u(t)$, which is commonly encountered in physical systems, e.g., with dipolar interactions.

For application to the orientation vector, one observes that only the component perpendicular to the orientation,
$\vec \omega_\perp(t) = \mathcal{Q}_{\vec u(t)} \, \vec \omega(t)$,
is relevant for the cross product in \cref{eq:kinematic-Ito};
it reads
\begin{multline}
  \vec \omega_\perp(t) =  (\zeta_R^\perp)^{-1} \mathcal{Q}_{\vec u(t)} \vec T_\ext(\vec u(t), t) \\
    + (\zeta_R^\perp)^{-1} \mathcal{Q}_{\vec u(t)} \vec \xi(t) \,,
  \label{eq:omega-perp}
\end{multline}
which contains only the friction coefficient $\zeta_R^\perp$.
Also, the evolution of $\vec \omega_\perp(t)$ does not depend on the remaining component $\vec \omega_\parallel(t) = \mathcal{P}_{\vec u(t)} \, \vec \omega(t)$, parallel to the particle axis.
This allows one to replace $\vec \omega(t)$ in \cref{eq:kinematic-Ito} by a process $\tilde{\vec \omega}(t)$, which
satisfies \cref{eq:euler-overdamped} for isotropic friction, i.e., to set $\zeta_R^\parallel = \zeta_R^\perp = \zeta_R$.
As an important consequence, the above discussion and analysis of the overdamped dynamics of $\vec u(t)$ for spherical particles carries over unchanged to axisymmetric particles. In particular, the orientation vector performs a homogeneous Brownian motion on the sphere (for time-independent $\vec T_\ext$) and the described numerical integration scheme can be employed without modifications upon choosing $\zeta_R = \zeta_R^\perp$.

The last point may also be seen from a different angle:
since the white noise $\vec \xi(t')$ and the orientation $\vec u(t)$ are independent for $t' \geq t$, the term $\mathcal{Q}_{\vec u(t)} \vec \xi(t)$ in \cref{eq:omega-perp} is equivalent (in probability) to an effectively two-dimensional, Gaussian white noise process that, for each time $t$, is rotated into the plane perpendicular to $\vec u(t)$.
This construction is already reflected in the algorithm when choosing the rotation angle in \cref{eq:Omega-components}.
We note that the algorithm is also applicable if $\vec T_\ext(\vec u, t)$ depends explicitly on time; however, the error estimates in \cref{sec:geometric} are no longer valid in this case.


For completeness, we specify the mean and covariance of $\vec \omega_\perp(t)$ for $t'\geq t$, conditioned on the value of $\vec u(t)$ [cf.\ \cref{eq:omega-moments}]:
\begin{subequations}
\begin{align}
  \langle \vec \omega_\perp(t) | \vec u(t) \rangle &= (\zeta_R^\perp)^{-1} \mathcal{Q}_{\vec u(t)} \vec T_\ext(\vec u(t), t) \,,
  \label{eq:omega-moments-anisotropic-mean} \\
  \langle \delta \vec \omega_\perp(t') \otimes \delta \vec \omega_\perp(t) | \vec u(t) \rangle &=
  2 D_R^\perp \mathcal{Q}_{\vec u(t)} \, \delta(t'-t) \,,
  \label{eq:omega-moments-anisotropic-cov}
\end{align}
\end{subequations}
where the condition is denoted on the right of the bar and
$\delta \vec \omega_\perp(t) := \vec \omega_\perp(t) - \langle \vec \omega_\perp(t) | \vec u(t) \rangle$.
In the second equation, the perpendicular part of the $\vec u(t)$-dependent rotational diffusion tensor enters:
\begin{equation}
  \tens D_R(t) := D_R^\parallel \mathcal{P}_{\vec u(t)} + D_R^\perp \mathcal{Q}_{\vec u(t)} \,, \\
\end{equation}
with constants $D_R^{\parallel,\perp} = \kB T / \zeta_R^{\parallel,\perp}$.
A derivation and further explanations of \cref{eq:omega-moments-anisotropic-mean,eq:omega-moments-anisotropic-cov} can be found in \cref{sec:omega-statistics}.


\section{Fokker--Planck equation}
\label{sec:FP-equation}

Before closing, we discuss the Fokker--Planck equation, which determines the propagator of the process
and which is complementary to the Langevin description of the stochastic evolution of the trajectories [\cref{eq:kinematic-Str,eq:kinematic-Ito}].
The (two-time) propagator $p(\vec u, t|\vec u_0, t_0)$ is the probability density of the orientation $\vec u(t)$ at time $t$ given its value $\vec u_0 = \vec u(t_0)$ at an earlier time $t_0 < t$.
Due to the Markov property of diffusion, $p(\vec u, t|\vec u_0, t_0)$ specifies the complete statistics of the dynamics (which is also referred to as the ``probability law'' of the process).
The Fokker--Planck equation is a partial differential equation (PDE) with an elliptic differential operator $\Lu(t)$ acting on the dependence on $\vec u$:
\begin{equation}
  \partial_t p(\vec u, t|\vec u_0, t_0) = \Lu(t) \, p(\vec u, t|\vec u_0, t_0)
  \label{eq:FPE}
\end{equation}
for $t\geq t_0$ and subject to the initial condition $p(\vec u, t_0|\vec u_0, t_0) = \delta(\vec u - \vec u_0)$.
For a time-homogeneous process, $\Lu$ is constant and the propagator depends only on $t - t_0$, so we can set $t_0=0$ and omit this dependence.
For free diffusion on a manifold,
$\Lu$ is the Laplace--Beltrami operator \cite{Oksendal:Stochastic,Kallenberg:Probability}, which in case of the unit sphere is simply the angular part of the three-dimensional Laplacian, see \cref{eq:Lu-diff-spherical} and Refs.~\cite{Coffey:LangevinEquation, BernePecora:DynamicLightScattering}.

\subsection{Fokker--Planck operator}

The link between the Langevin and Fokker--Planck pictures is straightforward for an Itō diffusion \cite{Gardiner:Stochastic,Oksendal:Stochastic}:
\begin{equation}
  \diff \vec X(t) = \vec b(\vec X(t), t) \, \diff t + \sigma(\vec X(t), t) \, \diff\vec W(t) \,,
  \label{eq:Ito-diffusion}
\end{equation}
where $\vec X(t)$ is vector-valued in $\mathbb{R}^d$, $\vec b(\cdot, t)$ is a $d$-dimensional, time-dependent vector field, and
$\vec W(t)$ is a standard Wiener process in $n$ dimensions scaled by the tensor field $\sigma(\cdot, t)$, which is $d\times n$-matrix-valued and also may (nicely) depend on time.
Then, the Fokker--Planck operator corresponding to \cref{eq:Ito-diffusion} reads
\begin{multline}
  \mathcal{L}_{\vec X}(t) f(\vec x) = -\nabla \cdot [\vec b(\vec x, t) f(\vec x)] \\
    + \frac{1}{2} \nabla \nabla^\top : \bigl[\sigma(\vec x, t) \, \sigma(\vec x, t)^\top f(\vec x) \bigr]
  \label{eq:FP-operator}
\end{multline}
for suitable real-valued functions~$f$; the nabla symbol $\nabla = (\partial_1, \dots, \partial_d)$ denotes the vectorial derivative w.r.t.\ the components of $\vec x$.

The form \eqref{eq:Ito-diffusion} fits the Itō SDE of rotational Brownian motion, \cref{eq:kinematic-Ito}, upon
setting $d=n=3$ and identifying $\vec X(t) = \vec u(t)$ and
$(2\DR)^{1/2} \, \diff \vec W(t) = \delta \vec \omega(t) \, \diff t$ [\cref{eq:omega-moments,eq:omega-moments-anisotropic-mean,eq:omega-moments-anisotropic-cov}],
see also the text after \cref{eq:Str-differential}.
To simplify the notation, we have dropped the superscript $\perp$ at the diffusion and friction coefficients.
One reads off that
\begin{align}
  b(\vec u, t) &= - \tauR^{-1} \vec u - \zeta_R^{-1} (\vec u \cdot \vec J) \, \vec T_\ext(\vec u, t) \,, \notag
\\
  \sigma(\vec u) &= -(\vec u \cdot \vec J) (2\DR)^{1/2} \,.
  \label{eq:FP-coefficients}
\end{align}
It remains to evaluate \cref{eq:FP-operator} for these choices of $\vec b$ and $\sigma$.
After a few calculations to favorably expose the divergence structure of $\Lu$ (see \cref{sec:FP-representations}),
one finds for the Fokker--Planck operator of rotational Brownian motion, corresponding to \cref{eq:kinematic-Ito}:
\begin{equation}
  \Lu f(\vec u) = -\nabla \cdot [\vec j(\vec u, t) f(\vec u)] \,,
  \label{eq:FP-rotation}
\end{equation}
in terms of the probability flux operator $\vec j(\vec u, t)$ separated into its drift and diffusive contributions [cf.\ \cref{eq:FP-rotation-app}]:
\begin{align}
  \vec j(\vec u, t) &:= \vec j_\text{drift}(\vec u, t) + \vec j_\text{diff}(\vec u) \,, \notag \\
  \vec j_\text{drift}(\vec u, t) &:=  -\zeta_R^{-1} (\vec u \cdot \vec J) \, \vec T_\ext(\vec u, t) \,, \notag \\
  \vec j_\text{diff}(\vec u) &:= -\DR |\vec u|^2 \mathcal{Q}_{\vec u} \nabla \,.
  \label{eq:FP-fluxes}
\end{align}
As above, $\mathcal{Q}_{\vec u} := \mathds{1} - |\vec u|^{-2} \, \vec u \otimes \vec u$
is the orthogonal projector onto the plane perpendicular to $\vec u$.

Normalization of $\vec u$ must not be assumed at this point since \cref{eq:FP-operator} is an unconstrained PDE problem in $\mathbb{R}^3$; rather it needs to be proven that the probability flux generated by $\Lu(t)$ preserves the norm of $\vec u$.
Both fluxes are obviously perpendicular to the orientation, $\vec u \cdot \vec j_\text{drift} = 0$ and $\vec u \cdot \vec j_\text{diff} = 0$, due to the cross product in the drift part and the projector $\mathcal{Q}_{\vec u}$ in the diffusive part.
We conclude that if $|\vec u_0|=1$ the solution $p(\vec u, t|\vec u_0, t_0)$ of \cref{eq:FPE} remains concentrated on the unit sphere under time evolution;
this holds also true if one uses a normalized initial distribution rather than a fixed value $\vec u_0$.
Under this condition, we can put $|\vec u|=1$ in \cref{eq:FP-rotation,eq:FP-fluxes} so that $\Lu(t)$ is actually a differential operator that is restricted to the unit sphere.
The latter fact is evident in a representation of $\Lu(t)$ in terms of spherical coordinates, which contains only derivatives w.r.t.\ the polar and azimuthal angles $(\theta,\phi)$, but not the norm $\rho=|\vec u|$ [\cref{eq:Lu-diff-spherical,eq:Lu-drift-spherical}].
Hence, the norm of $\vec u$ is conserved under the time evolution generated by $\Lu(t)$.

\subsection{Detailed balance}
\label{sec:detailed-balance}

We emphasize that, in the absence of the external torque, \cref{eq:FP-rotation} does not contain a drift term, $\vec j_\text{drift}=0$, and describes unbiased, free rotational diffusion.
The term $-\tauR^{-1} \vec u$ in $\vec b(\vec u,t)$ is merely an apparent drift and does not contribute to the probability flux, due to $\tauR = 2\DR$.
One verifies that $\Lu$ indeed coincides with the expected free diffusion operator on the sphere, i.e., the angular part of the Laplacian multiplied by $\DR$ [\cref{eq:Lu-diff-spherical}], which is derived in \cref{sec:FP-representations}.
It follows that the only stationary distribution, solving $\Lu \bar p(\vec u) = 0$, is the uniform distribution, $\bar p(\vec u) = 1 / 4\pi$, which also satisfies detailed balance, $\vec j_\text{diff}(\vec u) \, \bar p(\vec u) = 0$, and, hence, describes equilibrium.

More generally, if the external torque arises from a time-independent potential energy $V(\vec u)$ such that
\begin{equation}
  (\vec u \cdot \vec J) \vec T_\ext(\vec u) = |\vec u|^2 \mathcal{Q}_{\vec u} [\nabla V(\vec u)] \,,
  \label{eq:torque-potential}
\end{equation}
the following holds: The probability fluxes \eqref{eq:FP-fluxes} satisfy detailed balance,
\begin{equation}
  [\vec j_\text{drift}(\vec u) + \vec j_\text{diff}(\vec u)] \, p_\text{eq}(\vec u) = 0 \,,
  \label{eq:detailed-balance}
\end{equation}
if and only if the distribution $p_\text{eq}(\vec u)$ assumes the Boltzmann form,
\begin{equation}
  p_\text{eq}(\vec u) \propto \exp\boldsymbol({-\beta V(\vec u)}\boldsymbol) \,,
  \label{eq:boltzmann}
\end{equation}
characteristic of equilibrium at inverse temperature $\beta:=(\kB T)^{-1} = (\DR\zeta_R)^{-1}$.
Such $p_\text{eq}(\vec u)$ is trivially also a stationary solution, i.e., $\Lu p_\text{eq}(\vec u)=0$.

The statement is proven by, first, noting the equivalence of \cref{eq:detailed-balance} with
\begin{equation}
  \mathcal{Q}_{\vec u} \bigl\{ \beta [\nabla V(\vec u)] \, p_\text{eq}(\vec u) + \nabla p_\text{eq}(\vec u) \bigr\} = 0 \,.
  \label{eq:detailed-balance2}
\end{equation}
It is obvious that \cref{eq:boltzmann} satisfies this condition.
Conversely, searching for solutions to \cref{eq:detailed-balance2}, we exploit the positivity of the probability density and write $p_\text{eq}(\vec u) = \exp\boldsymbol({-\Phi(\vec u)}\boldsymbol)$, which implies that
\begin{equation}
  \mathcal{Q}_{\vec u} \bigl[ \beta \nabla V(\vec u) - \nabla \Phi(\vec u) \bigr] = 0 \,.
\end{equation}
Therefore, the term in brackets either vanishes or points parallel to the direction of $\vec u$.
It follows that
$\Phi(\vec u) = \beta V(\vec u) + \chi(\vec u)$
for some field $\chi(\cdot)$ with $\nabla \chi(\vec u) \parallel \vec u$.
Hence, $\chi(\vec u)$ depends only on the magnitude of $\vec u$, but not its direction, and reduces to a ($|\vec u|$-dependent) constant, which is fixed by the normalization of $p_\text{eq}(\vec u)$.
We conclude that \cref{eq:boltzmann} is the only solution to \cref{eq:detailed-balance}.

\subsection{Uniqueness of the stationary solution}
\label{sec:uniqueness}

Finally, we show that the equilibrium solution \eqref{eq:boltzmann} is the only stationary solution $\bar p(\vec u)$ if the torque is derived from a potential energy $V(\vec u)$, see \cref{eq:torque-potential}.
For any $\bar p(\vec u)$ with  $\Lu \bar p(\vec u) = 0$, we define $q(\vec u) = \exp\boldsymbol({\beta V(\vec u)}\boldsymbol) \, \bar p(\vec u)$.
Combining with \cref{eq:FP-rotation,eq:FP-fluxes,eq:torque-potential}, it implies
\begin{equation}
 \Lu \bar p(\vec u) = \nabla \cdot \bigl[ D_R \, \e^{-\beta V(\vec u)} |\vec u|^2 \mathcal{Q}_{\vec u}
  \nabla q(\vec u) \bigr] = 0 \,;
\end{equation}
also see \cref{eq:detailed-balance2} for the expression in brackets.
Without loss of generality, we can fix $|\vec u| = 1$.
Taking the $L^2$-inner product of $q(\vec u)$ and $\Lu \bar p(\vec u)$ on the unit sphere $S=\{ \vec u ; |\vec u|=1 \}$, we have
\begin{equation}
 \int_S q(\vec u) \, \nabla \cdot \bigl[ D_R \, \e^{-\beta V(\vec u)} \mathcal{Q}_{\vec u}
  \nabla q(\vec u) \bigr] \diff \vec u = 0 \,.
\end{equation}
By the generalized divergence theorem (Stokes' theorem for manifolds) \cite{Chirikjian:StochasticsLieGroups} and observing that the sphere has no boundary, it follows
\begin{equation}
 -\int_S \nabla q(\vec u) \, \cdot \bigl[ D_R \, \e^{-\beta V(\vec u)} \mathcal{Q}_{\vec u}
  \nabla q(\vec u) \bigr] \diff \vec u = 0 \,,
\end{equation}
which can be re-arranged, using $\DR > 0$ and the projection property $\mathcal{Q}_{\vec u}^2 = \mathcal{Q}_{\vec u}$, to read
\begin{equation}
 \int_S \e^{-\beta V(\vec u)} \bigl|\mathcal{Q}_{\vec u} \nabla q(\vec u) \bigr|^2 \, \diff \vec u = 0 \,.
\end{equation}
The integrand is non-negative, so it must vanish almost everywhere; in particular,
\begin{equation}
 \mathcal{Q}_{\vec u} \nabla q(\vec u) = 0 \quad \text{for all~} \vec u \in S \,,
\end{equation}
since $\nabla q(\vec u)$ is continuous.
In the same way as above for $\chi(\vec u)$, it follows that $\nabla q(\vec u) \parallel \vec u$ and that $q(\vec u)$ can depend only on the magnitude $|\vec u|$, which is fixed.
This is also seen from writing the operator $\mathcal{Q}_{\vec u} \nabla$ in spherical coordinates [\cref{eq:Q-nabla}].
Hence, $q(\vec u) = \text{const}$, and we conclude that \cref{eq:boltzmann} is the only stationary probability distribution in case of a potential torque.

\subsection{Example: dipole in a homogeneous field}

A typical application is the alignment of a magnetic dipole $\mu \vec u$ (with $|\vec u|=1$) in a spatially homogeneous and constant magnetic field $\vec B$; see e.g.\ Ref.~\cite{Boniface:2024} for an experiment.
The potential energy of the dipole is $V(\vec u) = -\mu \vec B \cdot \vec u$, which induces the torque
$\vec T_\ext(\vec u) = \mu \vec u \times \vec B$.
It results in [cf.\ \cref{eq:torque-potential}]
\begin{align}
  \vec u \times \vec T_\ext(\vec u) &= -\mu \vec B + \vec u (\vec u \cdot \mu\vec B) \notag \\
  &= \mathcal{Q}_{\vec u} \nabla V(\vec u)
\end{align}
and, thus, the equilibrium distribution [\cref{eq:boltzmann}]
\begin{equation}
  p_\text{eq}(\vec u) = \frac{\mu B/\kB T}{4\pi \sinh(\mu B/\kB T)} \, \exp(\mu\vec B\cdot \vec u / \kB T) \,.
  \label{eq:boltzmann-dipole}
\end{equation}
\Cref{sec:dipole} provides equivalent expressions of these results using spherical coordinates alongside with detailed calculations.

\section{Summary and conclusions}

Brownian motion on the unit sphere is a central ingredient to models of axisymmetric colloidal particles, in which it describes the stochastic evolution of the symmetry axis; examples are oblate and prolate ellipsoids, but also active Brownian particles, e.g., Janus spheres, where such an axis is given by the propulsion direction.
Here, we have constructed the corresponding overdamped stochastic process as a generalization of Brownian motion in flat (Euclidean) spaces, specifically, through a sequence of infinitesimal random rotations
[\cref{eq:inf_rotations}].
The latter construction is known as McKean--Gangolli injection \cite{Chirikjian:StochasticsLieGroups}, which transforms a stochastic process on the Lie algebra --- here, $\mathfrak{so}(3)$ --- into the generated Lie group --- here, the rotation group $SO(3)$.
Within Itō calculus, it is then straightforward to derive a stochastic differential (i.e., Langevin) equation of the process [\cref{eq:kinematic-Ito}].
Transforming the equation to its equivalent form in Stratonovich calculus [\cref{eq:kinematic-Str}] yields the same equation as is obtained heuristically from the underdamped Langevin equation of rotational Brownian motion [\cref{eq:kinematic,eq:euler-langevin}].
In particular, the latter form is free from the apparent, noise-induced drift term present in the Itō form. We have shown that, indeed, this additional term does not generate a drift (bias) of the rotational motion; but rather, it is needed to preserve the normalization of the orientation vector, $|\vec u(t)| = 1$, under the stochastic dynamics.
We have also derived the Fokker--Planck operator $\Lu$ corresponding to the overdamped Langevin equation [\cref{eq:FP-rotation,eq:FP-fluxes}]. Provided that the external torque originates from a potential energy [\cref{eq:torque-potential}], the form of $\Lu$ has allowed us to prove that the constructed process obeys detailed balance [\cref{eq:detailed-balance}] and uniquely yields the expected Boltzmann form of the equilibrium distribution [\cref{eq:boltzmann}], which is the only stationary distribution in this case (\cref{sec:uniqueness}).

In generic numerical integration schemes of the overdamped Langevin equation (in either form), such as the Euler--Maruyama and Milstein schemes, the normalization constraint is satisfied only asymptotically for vanishing time step, ${\Delta t \to 0}$, depending on their strong order of convergence.
As an alternative, we have proposed a geometric integration scheme [\cref{eq:finite_rotation}], which is based on finite random rotations and which satisfies the constraint exactly for any time step $\Delta t$.
Combining this approach with Gaussian distributed random rotation angles yields an immediate and simple implementation of this geometric integrator (\cref{tab:algorithm}), which we have shown to converge weakly at order 1, i.e., $O(\Delta t)$, and which can also be combined with a deterministic bias, e.g., due to an external torque or to model circle swimmers.
Moreover, we have outlined an improved algorithm that exactly generates the (free) propagator of rotational Brownian motion for an arbitrary time step $\Delta t$ if an advanced sampling of the rotation angle is used [see \cref{eq:theta-dist}].
Finally, we have demonstrated numerically and by deriving the corresponding propagators that the proposed Gaussian geometric scheme converges more rapidly than a widely used projection scheme [Eqs.~\eqref{eq:adhoc-scheme}].
The smaller prefactor of the discretization error allows for an integration time step that is an order of magnitude larger (\cref{fig:correlations});
in particular, the dynamics of free rotational diffusion is accurately generated also for time steps as large as $\DR \Delta t \approx 0.3$, i.e., typical rotation angles of $\theta \approx \pi/4$ in a single step (\cref{fig:propagators}).
As a consequence, suitable time steps of the geometric scheme are determined by the physical torques and not the constraint on $|\vec u|$.
Exact expressions for the orientational correlation functions [\cref{eq:corr-exact}] and for the equilibrium distribution of rotational Brownian motion [\cref{eq:boltzmann-dipole}] may serve as references to test actual implementations of the algorithm;
an exemplary implementation is provided in the open-source software \emph{HAL's MD package} \cite{HALMD} for hardware-accelerated simulation and analysis of many-particle systems.

\begin{acknowledgments}
 Financial support by Deutsche Forschungsgemeinschaft (DFG, German Research Foundation) under Germany’s Excellence Strategy---MATH+: The Berlin Mathematics Research Center (EXC-2046/1)---Project No.\ 390685689 (Subprojects AA1-18 and EF4-10) and further under Project No.\ 523950429 is gratefully acknowledged.
\end{acknowledgments}

\bigskip

\appendix

\section{Statistics of the projected angular velocity}
\label{sec:omega-statistics}

For a derivation of the first two moments of the projected angular velocity $\omega_\perp(t)$, given in \cref{eq:omega-moments-anisotropic-mean,eq:omega-moments-anisotropic-cov}, we switch to the integrated noise process,
\begin{equation}
  \vec \Omega_\perp(t) = \int_0^t \vec \omega_\perp(t') \, \diff t' \,,
\end{equation}
which in differential form reads $\diff \vec \Omega_\perp(t) = \vec \omega_\perp(t)\,\diff t$.
It is a proper Itō processes and, using \cref{eq:xi-moments,eq:omega-perp}, it obeys the following (Itō) SDE in standard form \cite{Gardiner:Stochastic,Oksendal:Stochastic}:
\begin{multline}
  \diff \vec \Omega_\perp(t) =  (\zeta_R^\perp)^{-1} \mathcal{Q}_{\vec u(t)} \vec T_\ext(\vec u(t), t) \, \diff t \\
    + (2 D_R^\perp)^{1/2} \mathcal{Q}_{\vec u(t)} \,\diff \vec W(t) \,,
  \label{eq:dOmega}
\end{multline}
which is driven by a standard three-dimensional Wiener process $\vec W(t)$ normalized such that $\langle|\vec W(t)|^2\rangle = 3 t$.

The mean value of a small increment $\Delta \vec \Omega_\perp(t) := \vec \Omega_\perp({t+\Delta t}) - \vec \Omega_\perp(t)$, conditioned on $\vec u(t)$, is given by the drift part only:
\begin{align}
  \mathrlap{E[\Delta \vec \Omega_\perp(t) | \vec u(t)]} \hspace{3em} \notag \\
  &= E\biggl[\int_t^{t+\Delta t} (\zeta_R^\perp)^{-1} \mathcal{Q}_{\vec u(t')} \vec T_\ext(\vec u(t'), t') \, \diff t' \biggl| \vec u(t) \biggr]  \notag \\
  &\simeq (\zeta_R^\perp)^{-1} \mathcal{Q}_{\vec u(t)} \vec T_\ext(\vec u(t), t) \Delta t \,; \quad \Delta t \to 0.
  \label{eq:Omega-perp-mean}
\end{align}
Here, $E[\,\cdot\,|\vec u(t)]$ denotes the conditional expectation \cite{Oksendal:Stochastic} with respect to the values of $\vec u(t)$ and, to obtain the second line, we have used continuity of the integrand.
The noise part of $\diff \vec \Omega_\perp(t)$ does not contribute since $\Delta \vec W(t) := \vec W(t+\Delta t) - \vec W(t)$ for $\Delta t > 0$ is independent of $\vec u(t)$ and thus:
\begin{equation}
  E\biggl[
    \int_{t}^{t'+\Delta t} 
    \mathcal{Q}_{\vec u(t')} \, \diff \vec W(t')
  \biggl| \vec u(t) \biggr]
  = 0.
\end{equation}
Switching to the derivative of $\vec \Omega_\perp(t)$ in \cref{eq:Omega-perp-mean} yields \cref{eq:omega-moments-anisotropic-mean}.

\bigskip

For the covariance function of $\vec \Omega_\perp(t)$, we compensate the drift and consider the rescaled process
$\diff \vec W^\perp(t) = \mathcal{Q}_{\vec u(t)} \, \diff \vec W(t)$ such that
$\int_0^t \delta \vec\omega_\perp(t') \,\diff t' = (2 D_R^\perp)^{1/2} \vec W^\perp(t)$.
Non-overlapping increments of $\vec W^\perp(t)$ are uncorrelated, which is seen by choosing $\Delta t', \Delta t> 0$ and $t' > t + \Delta t$
and performing standard manipulations of conditional expectation \cite{Oksendal:Stochastic}:
\begin{align}
  \mathrlap{E[\Delta W^\perp_i(t') \Delta W^\perp_j(t) | \vec u(t)]} \hspace{3em} \notag \\
   &= E[E[\Delta W^\perp_i(t') | \mathcal F(t')] \, \Delta W^\perp_j(t) | \vec u(t)] \\ 
   &= E[E[\Delta W^\perp_i(t') | \vec u(t')] \, \Delta W^\perp_j(t) | \vec u(t)] \notag \\   
   &= 0 \,, 
\end{align}
where we have used
the so-called tower property of $E[\cdot]$ in the first step, the Markov property of $\vec W^\perp(t)$ in the second,
and finally,
$E[\Delta W^\perp_i(t') | \vec u(t')] = E[\Delta W^\perp_i(t')] = 0$ due to the independence of $\Delta W^\perp_i(t')$ and $\vec u(t')$;
the symbol $\mathcal F(t)$ denotes the history of $\vec u$ up to time $t$.
Choosing the same time interval for both increments, it holds:
\begin{widetext}
\begin{align}
  E[\Delta W^\perp_i(t) \, \Delta W^\perp_j(t) | \vec u(t)]
  &= E\biggl[\int_t^{t+\Delta t} [\mathcal{Q}_{\vec u(t')}]_{ik} \diff W_k(t') \int_t^{t+\Delta t} [\mathcal{Q}_{\vec u(t'')}]_{jl} \diff W_l(t'')
  \biggl| \vec u(t) \biggr]  \notag \\
  &= E\biggl[\int_t^{t+\Delta t} [\mathcal{Q}_{\vec u(t')}]_{ik} [\mathcal{Q}_{\vec u(t')}]_{jk} \diff t'
  \biggl| \vec u(t) \biggr]  \notag \\
  &= E\biggl[\int_t^{t+\Delta t} [\mathcal{Q}_{\vec u(t')}]_{ij} \diff t' \biggl| \vec u(t) \biggr]  \notag \\
  &\simeq [\mathcal{Q}_{\vec u(t)}]_{ij} \Delta t \,; \quad \Delta t \to 0 \,.
  \label{eq:Omega-perp-cov}
\end{align}
\end{widetext}
Here, we have used (i) the properties of the Itō integral, in particular, that the quadratic variation of $\vec W(t)$ is
$\diff W_k(t) \cdot \diff W_l(t) = \delta_{kl} \diff t$ in mnemonic form,
(ii) that $\mathcal{Q}_{\vec u}$ is a symmetric matrix with $\mathcal{Q}_{\vec u}^2 = \mathcal{Q}_{\vec u}$,
and (iii) the continuity of the integrand.
Expressing $\vec W^\perp(t)$ in terms of $\delta \vec \omega_\perp(t)$, one sees that \cref{eq:Omega-perp-cov} corresponds to \cref{eq:omega-moments-anisotropic-cov}.

\section{Representations of the Fokker--Planck operator}
\label{sec:FP-representations}

Here, we provide details on the derivation of the Fokker--Planck operator in the divergence form given in \cref{eq:FP-rotation} and its representation in spherical coordinates.

First, one establishes that, for any $\vec u \in \mathbb{R}^3$,
\begin{align}
 \sigma(\vec u) \, \sigma(\vec u)^\top
  &= 2\DR (\vec u \cdot \vec J) (\vec u \cdot \vec J)^\top \notag \\
  &= 2\DR \bigl(|\vec u|^2 \mathds{1} - \vec u \otimes \vec u\bigr) \notag \\
  &= 2\DR |\vec u|^2 \mathcal{Q}_{\vec u} \,,
\end{align}
using \cref{eq:FP-coefficients}, the definition $(J_i)_{jk} = -\epsilon_{ijk}$, and standard rules of the $\epsilon$ symbol.
Next, by means of the identity
$\nabla \cdot (|\vec u|^2 \mathcal{Q}_{\vec u})^\top = -2\vec u$,
which reads in component notation:
\begin{equation}
  \partial_j \bigl(|\vec u|^2 \delta_{ij} - u_i u_j\bigr)
    = -2 u_i
\end{equation}
for fixed index $i$, one observes that
\begin{align}
  \partial_j \left[ \sigma_{ik}(\vec u) \, \sigma_{kj}(\vec u) f(\vec u) \right] \hspace{-6em} \notag \\
  &= \partial_j \left[ 2\DR \bigl(|\vec u|^2 \delta_{ij} - u_i u_j\bigr) f(\vec u) \right] \notag \\
  &= -4 \DR u_i f(\vec u)
    + \sigma_{ik}(\vec u) \, \sigma_{kj}(\vec u) \, \partial_j f(\vec u) \,.
\end{align}
Thus, the last term of \cref{eq:FP-operator} can be rewritten as
\begin{multline}
  \frac{1}{2} \nabla \nabla^\top : [\sigma(\vec u) \, \sigma(\vec u)^\top f(\vec u)] \\
  = - \nabla \cdot \left[ 2 \DR \vec u f(\vec u)
    + \frac{1}{2} \sigma(\vec u) \, \sigma(\vec u)^\top \nabla f(\vec u) \right] \,.
\end{multline}
Inserting the coefficient functions given in \cref{eq:FP-coefficients},
one arrives at
\begin{multline}
  \Lu f(\vec u) = \nabla \cdot [\zeta_R^{-1} (\vec u \cdot \vec J) \, \vec T_\ext(\vec u, t)  f(\vec u)] \\
    + \nabla \cdot \left[ \DR |\vec u|^2 \mathcal{Q}_{\vec u} \nabla f(\vec u) \right] ,
  \label{eq:FP-rotation-app}
\end{multline}
which is \cref{eq:FP-rotation} with \cref{eq:FP-fluxes}.
In particular, the apparent drift term $-\tauR^{-1} \vec u$ in $\vec b(\vec u,t)$ cancels out for $\tauR = 2\DR$.

A representation of $\Lu$ in spherical coordinates $(\rho, \theta, \phi)$
follows from the representation of the nabla operator \cite{Jackson:Electrodynamics},
\begin{equation}
  \nabla = \vec e_\rho \partial_\rho + \frac{1}{\rho} \, \vec e_\theta \partial_\theta
    + \frac{1}{\rho \sin(\theta)} \, \vec e_\phi \partial_\phi \,,
\end{equation}
with the orthonormal vectors $\vec e_\rho = \vec u / |\vec u|$, $\vec e_\theta = \partial_\theta \vec e_\rho$, and $\vec e_\phi = \cos(\theta)^{-1} \partial_\phi \vec e_\theta$ forming a local coordinate frame at point $\vec u$.
For the spherical coordinates, we use the convention $\rho \geq 0$, $\theta \in [0,\pi]$, and $\phi\in[-\pi,\pi)$ such that
\begin{equation}
  \vec u = \begin{pmatrix}
    \rho \sin(\theta) \cos(\phi) \\
    \rho \sin(\theta) \sin(\phi) \\
    \rho \cos(\theta)
  \end{pmatrix} \,.
\end{equation}
The projector on the plane perpendicular to $\vec u$ reads
$\mathcal{Q}_{\vec u} = \mathds{1} - \vec e_\rho \otimes \vec e_\rho$,
which immediately implies that the radial derivative $\partial_\rho$ drops out in
\begin{equation}
  \mathcal{Q}_{\vec u} \nabla = \frac{1}{\rho} \, \vec e_\theta \partial_\theta
    + \frac{1}{\rho \sin(\theta)} \, \vec e_\phi \partial_\phi \,.
  \label{eq:Q-nabla}
\end{equation}
Furthermore, we may duplicate $\mathcal{Q}_{\vec u} = \mathcal{Q}_{\vec u}^2$ in \cref{eq:FP-rotation-app} and let one factor $\mathcal{Q}_{\vec u}$ act to the left. A short calculation shows that also
\begin{equation}
  \nabla \mathcal{Q}_{\vec u} = \frac{1}{\rho} \vec e_\theta \partial_\theta
    + \frac{1}{\rho \sin(\theta)} \vec e_\phi \partial_\phi \,.
\end{equation}
Hence, the diffusion part of $\Lu$ in \cref{eq:FP-rotation-app} is given by
\begin{align}
  \Lu^\text{diff} &= \nabla \cdot \DR |\vec u|^2 \mathcal{Q}_{\vec u} \nabla \notag \\
    &= \DR \Bigl(\vec e_\theta \partial_\theta
    + \frac{1}{\sin(\theta)} \, \vec e_\phi \partial_\phi \Bigr)^2 \notag \\
    &= \DR \left( \frac{1}{\sin(\theta)} \, \partial_\theta \sin(\theta) \, \partial_\theta
    + \frac{1}{\sin(\theta)^2} \, \partial_\phi^2 \right) \,,
  \label{eq:Lu-diff-spherical}
\end{align}
which is the angular part of the Laplacian in $\mathbb{R}^3$, as claimed in the main text.

Finally, for the drift part of $\Lu$, we note the identity
\begin{multline}
  \nabla \cdot \left[(\vec u \times \vec T_\ext) f \right] = \\
    - \frac{1}{\sin(\theta)} \, \partial_\theta [\sin(\theta) T_\phi f]
    + \frac{1}{\sin(\theta)} \, \partial_\phi (T_\theta f)
\end{multline}
in terms of the coefficients
$T_\alpha(\rho, \theta, \phi, t) = \vec e_\alpha \cdot \vec T_\ext(\vec u, t)$ of the external torque ($\alpha=\rho, \theta, \phi$) and some function $f(\vec u) = f(\rho,\theta,\phi)$.
It yields
\begin{align}
  \Lu^\text{drift} f(\vec u) &= \nabla \cdot \zeta_R^{-1} (\vec u \cdot \vec J) \, \vec T_\ext(\vec u, t) f(\vec u) \notag \\
  &= - \frac{\zeta_R^{-1}}{\sin(\theta)} \,\Bigl\{ \partial_\theta \bigl[\sin(\theta) T_\phi f\bigr]
     - \partial_\phi (T_\theta f)  \Bigr\} \,.
  \label{eq:Lu-drift-spherical}
\end{align}
In particular, $\Lu^\text{drift}$ is independent of $\rho$ and does not contain a derivative w.r.t.~$\rho$.

The equilibrium solution $p_\text{eq}(\vec u)$, in the case of a time-independent external torque,
is characterized by the detailed balance condition, $\vec j(\vec u) \, p_\text{eq}(\vec u) = 0$.
Using \cref{eq:FP-fluxes}, this condition reads explicitly
\begin{multline}
  \zeta_R^{-1} (\vec u \cdot \vec J) \, \vec T_\ext(\vec u) \, p_\text{eq}(\vec u) \\
  + \DR |\vec u|^2 \mathcal{Q}_{\vec u} \nabla p_\text{eq}(\vec u) = 0 \,;
\end{multline}
in spherical coordinates, it assumes the form
\begin{multline}
  \bigl[-\beta T_\phi(\theta, \phi) \, \vec e_\theta + \beta T_\theta(\theta, \phi) \, \vec e_\phi \bigr] p_\text{eq}(\theta,\phi)
    \notag \\
  + \Bigl[\vec e_\theta \partial_\theta + \frac{1}{\sin(\theta)} \, \vec e_\phi \partial_\phi\Bigr]
  p_\text{eq}(\theta, \phi) = 0 \,.
\end{multline}
This vector equation implies a system of two scalar equations, one for each of the independent directions $\vec e_\theta$ and $\vec e_\phi$:
\begin{align}
  \bigl[-\beta T_\phi(\theta, \phi) \, + \partial_\theta \bigr]
    p_\text{eq}(\theta, \phi) &= 0 \,, \notag \\
  \bigl[\sin(\theta)\, \beta T_\theta(\theta, \phi) + \partial_\phi\bigr]
    p_\text{eq}(\theta, \phi) &= 0 \,.
  \label{eq:detailed-balance-spherical}
\end{align}
This system of coupled differential equations is separable if $\partial_\phi T_\phi = 0$ and
$\partial_\theta[\sin(\theta) T_\theta]=0$; otherwise, it constitutes a genuine PDE problem.

\section{Dipole in a homogeneous field, using spherical coordinates}
\label{sec:dipole}

As an illustrative example, we consider a magnetic dipole $\mu\vec u$ in a homogeneous and constant magnetic field $\vec B$ and solve for the stationary probability distribution $\bar p(\vec u)$ of the Fokker--Planck equation \eqref{eq:FP-rotation}, which obeys $\Lu \bar p(\vec u) = 0$.
The generated torque, $\vec T_\ext(\vec u) = \mu \vec u \times \vec B$, tends to align the orientation $\vec u$ with the field $\vec B$.
Choosing the Cartesian frame such that $\vec B = B \vec e_3$ and switching to the spherical coordinates of
\cref{sec:FP-representations}, it follows
$T_\theta = 0$ and $T_\phi = -\mu B \sin(\theta)$.

\paragraph*{Equilibrium solution.}
Before we address the general stationary problem, we restrict to the equilibrium solution $p_\text{eq}(\theta,\phi)$, which obeys \cref{eq:detailed-balance-spherical}.
Inserting the concrete example and changing the variable $\theta$ to $z:=\cos(\theta) \in [-1,1]$
yields
\begin{align}
  (-\beta \mu B + \partial_z) \, p_\text{eq}(z, \phi) &= 0 , \notag \\
  \partial_\phi \, p_\text{eq}(z, \phi) &= 0 \,,
\end{align}
where we have used that $\sin(\theta)^{-1} \partial_\theta = - \partial_z$.
It has the solution
$
  p_\text{eq}(z, \phi) \propto \exp(\beta\mu B z),
$ 
or, after normalization and in vector notation:
\begin{equation}
  p_\text{eq}(\vec u) = \frac{\beta\mu B}{4\pi \sinh(\beta\mu B)} \, \exp(\beta\mu \vec B \cdot \vec u) \,.
  \label{eq:boltzmann-dipole-app}
\end{equation}

\paragraph*{Stationary solution.}
It remains to show that the equilibrium solution is the only stationary solution. To this end, we find the general solution to $\Lu \bar p(\vec u)=0$ for the present example.
Starting from \cref{eq:Lu-diff-spherical,eq:Lu-drift-spherical}, inserting $T_\theta$ and $T_\phi$, and multiplying by
$\sin(\theta)^2 / \DR$ yields the separable boundary value problem,
\begin{multline}
  \beta \mu B \sin(\theta)^2 [2\cos(\theta) + \sin(\theta) \partial_\theta] \, \bar p(\theta, \phi) \\
    + [\sin(\theta) \partial_\theta]^2 \, \bar p(\theta, \phi)
    + \partial_\phi^2  \, \bar p(\theta, \phi) = 0 \,,
  \label{eq:FP-dipole}
\end{multline}
with $\beta=(\DR\zeta_R)^{-1}$ and boundary conditions:
\begin{subequations}
\begin{align}
  \bar p(\theta, -\pi) &= \bar p(\theta, \pi) && \text{for all} \: \theta\in[0,\pi] \,,
  \label{eq:bc-phi} \\
  \partial_\phi \bar p(0, \phi) &= \partial_\phi \bar p(\pi, \phi) = 0 && \text{for all} \: \phi\in[-\pi,\pi] \,,
  \label{eq:bc-theta}
\end{align}
\end{subequations}
to ensure (a) periodicity in $\phi$ and
(b) differentiability at the poles, $|\nabla \bar p(\vec u)| < \infty$ for $\vec u=\pm \vec e_3$.
The problem is solved by the product ansatz $\bar p(\theta, \phi) = g\boldsymbol({z(\theta)}\boldsymbol) \, h(\phi)$,
with $z(\theta) = \cos(\theta)$ as above,
which leads to the coupled system of ordinary differential equations for the separation constant $\lambda$:
\begin{subequations}
\begin{align}
  \partial_z (1-z^2) (-\beta \mu B +\partial_z) \, g(z) &= \frac{\lambda g(z)}{1-z^2} \,,
  \label{eq:ode-g} \\
  -\partial_\phi^2 h(\phi) &= \lambda h(\phi) \,,
  \label{eq:ode-h}
\end{align}
\end{subequations}
using $\sin(\theta) \, \partial_\theta = -(1-z^2) \, \partial_z$.
\Cref{eq:ode-h} with the condition $h(-\pi) = h(\pi)$, see \cref{eq:bc-phi}, has non-zero solutions only for the eigenvalues $\lambda=m^2$ with integer $m=0,1,2,\dots$
The second boundary condition, \cref{eq:bc-theta}, reduces to
\begin{equation}
  \partial_\phi h(\phi) = 0 \quad \text{or} \quad g(\pm 1) = 0 \,.
  \label{eq:bc-theta-alt}
\end{equation}
Recalling that $\bar p(\theta, \phi)$ is a solution to the Fokker--Planck equation \eqref{eq:FP-dipole} and a probability density, we search for functions $g(z)$ and $h(\phi)$ which are twice continuously differentiable, non-negative, and normalizable; in particular, they are strictly positive on sub-intervals of $[-1,1]$.

The first case of \cref{eq:bc-theta-alt} selects $m=0$ and yields $h(\phi)=\text{const}$.
Integration of \cref{eq:ode-g} w.r.t.\ $z$ for $\lambda=0$ yields for some constant $c$:
\begin{equation}
  (-\beta \mu B +\partial_z) \, g(z) = \frac{c}{1-z^2} \,.
\end{equation}
Since $\partial_z g(z)$ is continuous, the l.h.s.\ of this equation is continuous for $z \in [-1,1]$ and, thus, bounded in magnitude.
Balancing with the r.h.s.\ requires that $c=0$ so that the divergence upon $z\to \pm 1$ is removed.
The remaining equation is homogeneous in $g(z)$ and its solutions are of the form
$g(z) \propto \exp(\beta \mu B z)$.

\paragraph*{Uniqueness.}
We use the obtained solution for $g(z)$ as an integrating factor to test for possible further solutions to \cref{eq:ode-g} if $\lambda = m^2 \geq 1$, amended by the second boundary condition in \cref{eq:bc-theta-alt}.
Defining $q(z)$ such that $g(z) =: \exp(\beta \mu B z) q(z)$, it holds
\begin{equation}
 (-\beta \mu B +\partial_z) \, g(z)  = \e^{\beta \mu B z} q'(z) \,.
\end{equation}
Inserting in \cref{eq:ode-g}, multiplying by $q(z)$, and integrating over $z\in[-1,1]$ yields
\begin{multline}
  \int_{-1}^1 q(z) \, \partial_z (1-z^2) \, \e^{\beta \mu B z} q'(z) \, \diff z \\
    = m^2 \int_{-1}^1 \frac{\e^{\beta \mu B z} q(z)^2}{1-z^2} \, \diff z > 0 \,.
  \label{eq:quadratic-form}
\end{multline}
The r.h.s.\ is strictly positive, since its integrand is non-negative and $q(z) > 0$ on a subinterval of $[-1,1]$.
But the l.h.s.\ cannot be positive, which is seen after integration by parts:
\begin{equation}
  \text{l.h.s.} = -\int_{-1}^1 \, (1-z^2) \, \e^{\beta \mu B z} [q'(z)]^2 \, \diff z \leq 0 \,;
  \label{eq:quadratic-form-lhs}
\end{equation}
the boundary terms vanish due to $q(\pm 1) = \e^{\mp \beta\mu B} g(\pm 1) = 0$ and since $|q'(z)|$ is bounded on $z\in [-1,1]$.
Therefore, \cref{eq:quadratic-form} cannot be satisfied by any (non-zero) $q(z)$.

A similar reasoning can also be applied for $m=0$, which enforces the l.h.s.\ of \cref{eq:quadratic-form} to be zero
and the integrand in \cref{eq:quadratic-form-lhs}, being non-negative, to vanish almost everywhere.
It follows that $q'(z) = 0$ for all $z$ and, thus, $q(z) = \text{const}$, recovering the above result for $g(z)$.

We conclude that the case $m=0$ yields the only stationary solution,
\begin{equation}
  \bar p(\theta, \phi) \propto \exp\boldsymbol({\beta \mu B \cos(\theta)}\boldsymbol) \,,
  \label{eq:boltzmann-dipole-spherical}
\end{equation}
which, of course, agrees with the previously obtained equilibrium solution $p_\text{eq}(\vec u)$, given in \cref{eq:boltzmann-dipole-app}, and which is compatible with the azimuthal symmetry of the potential generating the external torque.


\bibliography{rotation-note}

\end{document}